\documentclass[preprint,12pt,authoryear]{elsarticle}
\usepackage{booktabs}
\usepackage{amssymb}
\usepackage{amsmath}
\usepackage{natbib}
\usepackage{graphicx}
\usepackage[colorlinks=true,linkcolor=blue,citecolor=blue]{hyperref}

\def\apj{{ ApJ}}
\def\apjl{{ApJL}}

\def\aap{{ A\&A}}

\def\mnras{{ MNRAS}}

\def\nat {{ Nature}}
\def\pasj{{ PASJ}}

\def\physrep{{ Physics Reports}}

\def\prd{{ Phys. Rev. D}}
\def\PRL{{ Phys. Rev. Lett}}

\def\msun{\mr{M_{\odot}}}
\def\mp{m_{\rm p}}
\def\me{m_{\rm e}}
\def\epse{\epsilon_{\rm e}}
\def\epsB{\epsilon_{\rm B}}
\def\mr{\mathrm}

\def\ga{\gamma_{\rm a}}
\def\gm{\gamma_{\rm m}}

\def\d{\mathrm{d}}

\begin{document}
\begin{frontmatter}


\title{Jets from Tidal Disruption Events}

\author[1]{Fabio De Colle\corref{cor1}%
\fnref{fn1}}
\ead{fabio@nucleares.unam.mx}
\address[1]{Instituto de Ciencias Nucleares, Universidad Nacional Aut\'onoma de M\'exico, A. P. 70-543 04510 D. F. Mexico}

\author[2]{Wenbin Lu\fnref{fn2}}
\ead{wenbinlu@caltech.edu}
\address[2]{TAPIR, Walter Burke Institute for Theoretical Physics, Mail Code 350-17, Caltech, Pasadena, CA 91125, USA}


\begin{abstract}
The discovery of jets from tidal disruption events (TDEs) rejuvenated the old field of relativistic jets powered by accretion onto supermassive black holes. In this Chapter, we first review the extensive multi-wavelength observations of jetted TDEs. Then, we show that these events provide valuable information on many aspects of jet physics from a new prospective, including the on-and-off switch of jet launching, jet propagation through the ambient medium, $\gamma/$X-ray radiation mechanism, jet composition, and the multi-messenger picture. Finally, open questions and future prospects in this field are summarized.

\end{abstract}


\end{frontmatter}

\section{Introduction}

Most if not all massive galaxies host a super-massive black hole (SMBH) at their center. 
In active galactic nuclei (AGN), SMBHs eject relativistic jets with typical Lorentz factors of $\sim 10$, which produce luminous multi-wavelength emission ranging from radio to gamma rays. Nevertheless, most SMBHs have long quiescent times with short periods of activity (once every $\sim 10^4$~yrs, e.g., \citealt{2018ApJ...852...72V}) driven by tidal disruption events (TDEs), i.e. stars which come close enough to the SMBH to be ripped apart by tidal forces.

About $\sim 90$ tidal disruption event (TDE) candidates have been discovered so far \citep[see, e.g.,][for a review]{komossa15}\footnote{ https://tde.space/ presents an up to date catalog of observed TDEs}. Predicted by several authors \citep[e.g.,][]{hills75, 1982ApJ...262..120L, rees88,evans89}, they were discovered by the \textit{ROSAT} telescope in the late nineties \citep[e.g.,][]{bade96,grupe99,komossa99}. When detected in X-rays, their  outbursts are typically soft ($\lesssim 0.1$~keV) and very luminous ($\approx 10^{42}-10^{45}$~erg s$^{-1}$). While the rate the disrupted star accretes on the SMBH after the first close-by passage is predicted to drop with time approximately as $\sim t^{-5/3}$ \citep{evans89, 1989IAUS..136..543P}, observations show a shallower decay \citep{auchettl17,auchettl18}. More recently, many TDE candidates are found by the increasingly powerful optical/UV transient surveys \citep[e.g.][]{2008ApJ...676..944G, 2009ApJ...698.1367G, vanvelzen11, 2012Natur.485..217G, arcavi14, 2015ApJ...798...12V, holoien16}. Looking into the future, the optical-UV TDE sample size will increase by a factor of 10--100 given the unprecedented volume survey speed of \textit{ZTF} and \textit{LSST} \citep[e.g.,][]{2018arXiv180902608V}, and the \textit{eROSITA} all-sky survey is expected to discover a few thousand TDEs in the soft X-ray band \citep{2014MNRAS.437..327K}.

\begin{figure}
\centering
\includegraphics[scale=0.29]{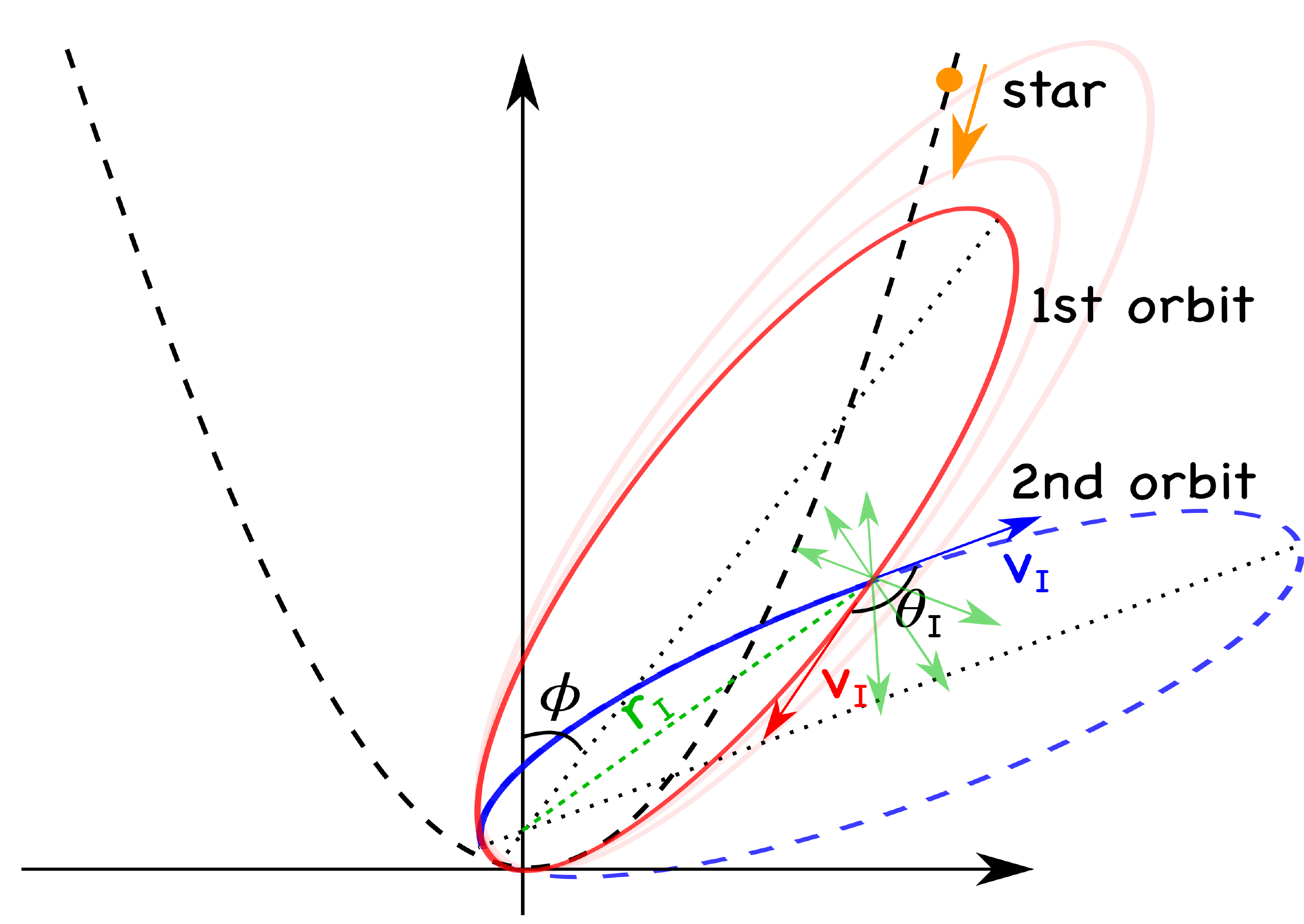}
\caption{The star was initially in a parabolic orbit (orange curve). After the tidal disruption, the bound materials are in highly eccentric elliptical orbits of different semimajor axes (red curves) but have nearly the same apsidal precession angle per orbital cycle. Materials in their second orbits (blue curves) collide with what is still in the first orbit. A fraction of the shocked gas may be ejected to infinity and the rest stays bound and form an accretion disk around the SMBH. The subject of this Chapter is on the relativistic jets launched by the accretion disk.
}\label{fig:orbit}
\end{figure}

The hydrodynamics of TDEs is complex and not fully understood \citep[see, e.g.,][for a recent review]{2018arXiv180110180S}. Initially, the star approaches the SMBH in a parabolic orbit (see Figure \ref{fig:orbit}). At the tidal radius $r_t = r_\star (M_{\rm BH}/M_\star)^{1/3}\simeq 7\times 10^{12} r_{\star,\odot} (M_{\rm BH,6}/M_{\star,\odot})^{1/3}$ (where the suffixes $_{\rm BH}$ and $_\star$ indicate the SMBH and the disrupted star respectively),
the star is shredded by the BH's tidal forces. About half of the stellar debris are in highly eccentric bound orbits falling back on timescales of $\sim$ months (for a solar-like main-sequence star), and the other half is ejected back to infinity. Numerical simulations show that the fallback stream dissipates its kinetic energy due to self-crossing shocks, which eventually lead to the formation of an accretion disk with a typical size $\sim r_t$ \citep{2015ApJ...804...85S, 2016MNRAS.455.2253B}. The temperature of the accreting material is expected to be $\sim$ a few times 10$^5$~K, so the expected peak frequency should fall in the undetectable EUV band \citep{ulmer99, strubbe09, 2011MNRAS.410..359L}. This may cause the majority of the radiative energy to be missing by current observations \citep{lu18}.

Since the typical fallback rate stays super-Eddington for $\sim$ a year
\citep[e.g.,][and references therein]{guillochon13}, radiation driven outflows or jets could be associated to the accretion process \citep[e.g.,][]{giannios11,decolle12}.
Since 2011, three jetted TDE candidates were discovered by $\gamma$-ray triggers from the \textit{Neil Gehrels Swift Observatory}. The extremely bright and variable $\gamma$/X-ray emission requires a strongly anisotropic radiation pattern generated by a relativistic jet pointing towards the observer \citep[e.g.,][]{bloom11, burrows11,zauderer11,cenko12,brown15}. Radio to mm-band afterglow emission follows the prompt $\gamma$/X-rays as the jet interacts with and is decelerated by the surrounding medium on timescales of months to years. Additionally, a few other TDE candidates have outflows or (off-axis) jets inferred from their radio afterglow \citep[e.g.,][]{alexander16, vanvelzen16, alexander17, saxton17,  mattila18}.

Jetted TDEs are an ideal testbed to study the jet physics. The three major puzzles of relativistic jets are: the jet launching process (what parameters control the on-and-off switch), the jet emission mechanism (what mechanisms convert the jet power into X/$\gamma$-ray radiation at high efficiency), and the jet composition (the energy fraction carried by baryons, leptons, and magnetic fields). As will be shown in this Chapter, jetted TDEs may provide answers to all three questions, thanks to the reasonably well-estimated accretion rate and total energy budget.

In addition, while AGN jets persist for long enough to dig a hole through the galaxy, TDE jets, created in otherwise quiescent galaxies, allow us to study the poorly known environment of SMBHs by the afterglow associated with the jet deceleration. 

We present here a review of jets in TDEs. In section \ref{sec:obs} we will describe existing observations of TDE jet candidates, while in Section \ref{sec:models} we will describe the theoretical implications of these observations. In Section \ref{sec:multi} we will brief review existing studies of multi-messenger astrophysics with TDE jets, and we will drive our conclusions in Section \ref{sec:concl}.

\section{Observations}
\label{sec:obs}

Seven candidate jet/outflows from TDEs have been observed so far (see Table 1).
Three of them have been discovered by \textit{Swift}, one by \textit{INTEGRAL}, while in the other cases the presence of fast moving material is inferred from radio emission. In this section, we briefly review existing observations. We first focus on the \textit{Swift} events (section \ref{sec:swift}), then on the others mainly characterized by their non-thermal radio emission (section \ref{sec:radio}). Finally, we discuss the implications of these observations on the rate of jets in TDEs (section \ref{sec:rate}).

\begin{table}
\label{tab:jets}
\begin{tabular}{l l l l l l}
\toprule
 Name &year& z & X-rays? &     \\  \midrule
 Swift J164449.3+573451  &2011/2011& 0.3543 & non-thermal  & On-axis jet \\
Swift J2058.4+0516  &2011/2012& 1.1853 & non-thermal & On-axis jet  \\
Swift J1112.2-8238  &2011/2015& 0.89  & non-thermal & On-axis jet  \\
\midrule
Arp 299-B AT1 & 2005/2018 &0.0103 & - & Off-axis jet\\ 
\midrule
ASASSN-14li &2014/2016& 0.0206 & thermal & Non-rel. outflow\\
XMMSL1 J0740-85 &2014/2017& 0.0173  & thermal & Non-rel. outflow \\ 
\midrule
 IGR J1258+0134 &2011/2015& 0.0033  & non-thermal & AGN? \\ 
\bottomrule
\end{tabular}
\caption{Candidate jet and/or outflows discovered so far. The columns indicate the year of the event/publication, the redshift, the presence of hard/soft X-rays and the possible interpretation of the observations (the references are reported in the text).}
\end{table}

\subsection{\textit{Swift} events}
\label{sec:swift}

The jetted TDE candidate \textit{Swift} J164449.3+573451 (hereafter SW1644) was detected by the \textit{Swift Burst Alert Telescope} (BAT) on 28 March 2011. Shortly after the trigger, multi-wavelength observations were performed from ground-based telescopes and from space observatories. Optical and near-IR observations showed a counterpart consistent with the position of SW1644, with H$\beta$ and [OIII] emission lines corresponding to a redshift of $z = 0.35$ \citep{barres11, bloom11, burrows11, levan11, zauderer11}. 

In the years after its discovery, SW1644 has been extensively observed in radio \citep{zauderer11, berger12, zauderer13, cendes14, yang16, wiersema12, eftekhari18}, infrared/optical \citep{levan16, wiersema12}
and X-rays \citep{saxton12, reis12, zauderer13, mangano16, eftekhari18, levan16}. Upper limits were also provided by \textit{VERITAS} \citep{aliu11}, \textit{MAGIC} \citep{aleksic13} and \textit{Fermi} LAT \citep{omodei11, peng16} above $\sim$~50 MeV.

The spatial coincidence between the radio emission and the center of the associated galaxy (within 150 pc, see \citealt{zauderer11}), and the decay of the X-ray flux as $\sim t^{-5/3}$ were strong indications of the association with a TDE \citep[but see][for a different interpretation]{quataert12}.

\begin{figure}[htbp!]
\centering
\includegraphics[scale=0.35]{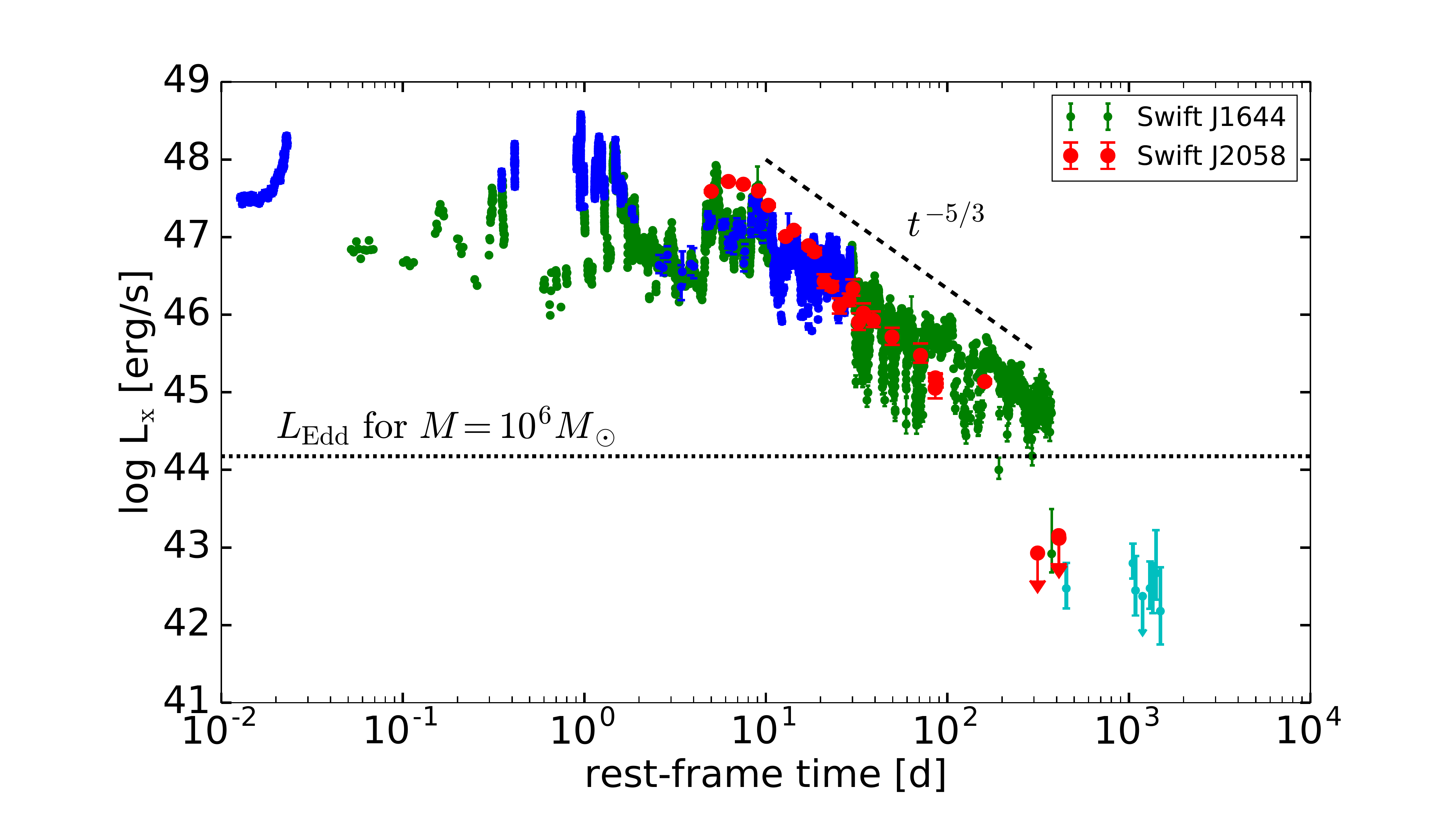}
\caption{X-ray lightcurves for the two jetted TDEs SW1644 (blue, green and cyan are from \textit{Swift} XRT PC mode, XRT WT mode, and \textit{Chandra}) and SW2058 (red).}
\label{fig:X-ray_LC}
\end{figure}

The X-ray light curve (see Figure \ref{fig:X-ray_LC}) is unlike any other X-ray transient discovered previously. During the first two weeks, it presented a series of flares with fluxes varying from $\sim 10^{45}$ to $\sim 10^{48}$~erg~s$^{-1}$ on timescales $\gtrsim 100$~s \citep{bloom11, burrows11}. The peak X-ray luminosity is a factor of $\sim10^4$ higher than the Eddington luminosity $L_{\rm Edd}$ of a $10^6M_\odot$ BH. The shortest flux-doubling time (in the host-galaxy rest frame) is a factor of $\sim$ 6 shorter than the redshifted period of the innermost stable circular orbit (ISCO) for a non-spinning $10^6 M_\odot$ BH. Variability on even shorter timescales is possible but not well constrained due to low photon statistics.

The total isotropic energy radiated in the X-ray band is $E_{\rm X}\sim 5\times 10^{53}\,$erg \citep{burrows11}. This is similar to the isotropic energy of a typical gamma-ray burst (GRB), but emitted during a much larger timescale $\sim$~10$^6$~s (vs. $\lesssim 10^4$~s even for ultra-long GRBs, see \citealt{levan14}). If one (conservatively) accounts for bolometric correction of a factor of $\eta_{\rm bol}\sim 1/3$ and radiative efficiency of another factor of $\eta_{\rm rad}\sim 1/3$, then the total isotropic energy of the jet is $E_{\rm j}\sim 5\times 10^{54} /(10 \; \eta_{\rm bol} \eta_{\rm rad})\,$erg. Comparing this with the total kinetic energy indicated by the late-time radio afterglow $E_{\rm K}\sim10^{53}\,$erg (see discussion in \S 3.2), we see that the jet must have a small beaming factor $f_{\rm b}\equiv \Omega_{\rm j}/4\pi\lesssim 0.02$ (only a fraction of $E_{\rm K}$ comes from the X-ray emitting jet). This, along with the rapid variability and hyper-Eddington X-ray luminosity, supports the picture of a relativistic jet pointing towards the observer like the case of blazars.

During the following months the X-ray emission dropped roughly as $t^{-5/3}$ (but see \citealt{mangano16}, who found the data being more consistent with a slightly shallower $t^{-1.5}$ decay). Up to $\sim 100$~days after the trigger, it still presented flares with flux increasing by an order of magnitude, then it entered into a plateau phase without strong flares until $t\sim1$~yr. At $\sim 500$~days, the flux dramatically drops by a factor of 100 becoming barely detectable by \textit{Swift} \citep{zauderer13, mangano16}. This also corresponds to the time when the mass fallback rate became smaller than the Eddington accretion rate $\sim 10 L_{\rm Edd}/c^2$ for a $\sim$ 10$^6$~M$_\odot$ BH, indicating a relation between super-Eddington accretion and jet ejection \citep[see, e.g.][]{decolle12, tchekhovskoy14}. Later observations (up to $t\sim7$~yr) showed a faint persistent X-ray source consistent with emission from the forward shock \citep{eftekhari18}. Alternatively, this emission could be due to the Compton echo of X-ray photons
of the transient source on surrounding electrons  \citep{cheng16} or to a low-luminosity AGN activity from the source \citep{levan16}.

Some periodicity has been detected in the X-ray lightcurve. Extended observations by the \textit{Suzaku} and \textit{XMM-Newton} observatories showed a quasi-periodic oscillation (QPO) of about $\sim$~200~s, possibly corresponding to the keplerian frequency at the innermost stable circular orbit (ISCO) of a BH with a mass in the range $4.5\times10^5$--$5 \times10^6$~M$_\odot$ \citep[for a non-rotating and maximumly rotating BH respectively,][]{reis12}. \citet{abramowicz12}, on the other hand, interpreted this QPO as due to an epicyclic frequency in the accretion disk, and in this case the BH mass is $\sim 10^5M_\odot$ (which allows the disruption of a white dwarf, as argued by \citealt{krolik11}).

\citet{saxton12} found that the low-flux states in the lightcurve are separated in several cases by $2.2\times10^5$~s (or multiples of this period). They interpreted it as evidence of jet precession, with flares/dips determined by the jet orientation with respect to the line of sight. This timescale is consistent with the Lense-Thirring precession period of a ring at radius $\sim$ 20 $r_{\rm g}$ (or 10 $r_{\rm g}$) from a BH of mass $10^6M_{\odot}$ (or $10^7 M_{\odot}$), where $r_{\rm g}=GM/c^2$ is the gravitational radius. This is also consistent with the source being harder when brighter. On the other hand, \citet{stone12} and \citet{tchekhovskoy14} pointed out that, for a narrowly beamed precessing jet with a large misalignment, the observer line of sight is only in the beaming cone for a very small fraction of the time $\theta_{\rm j}/2\pi\sim 10^{-2}$ (for a beaming angle $\theta_{\rm j}\sim 0.1\rm\,rad$). Thus, the persistent X-ray emission indicates that the jet is always very nearly aligned with our line of sight, perhaps due to the alignment torque by the magnetic flux accumulated on the BH \citep{tchekhovskoy14}.

\begin{figure}[htbp!]
\centering
\includegraphics[scale=0.5]{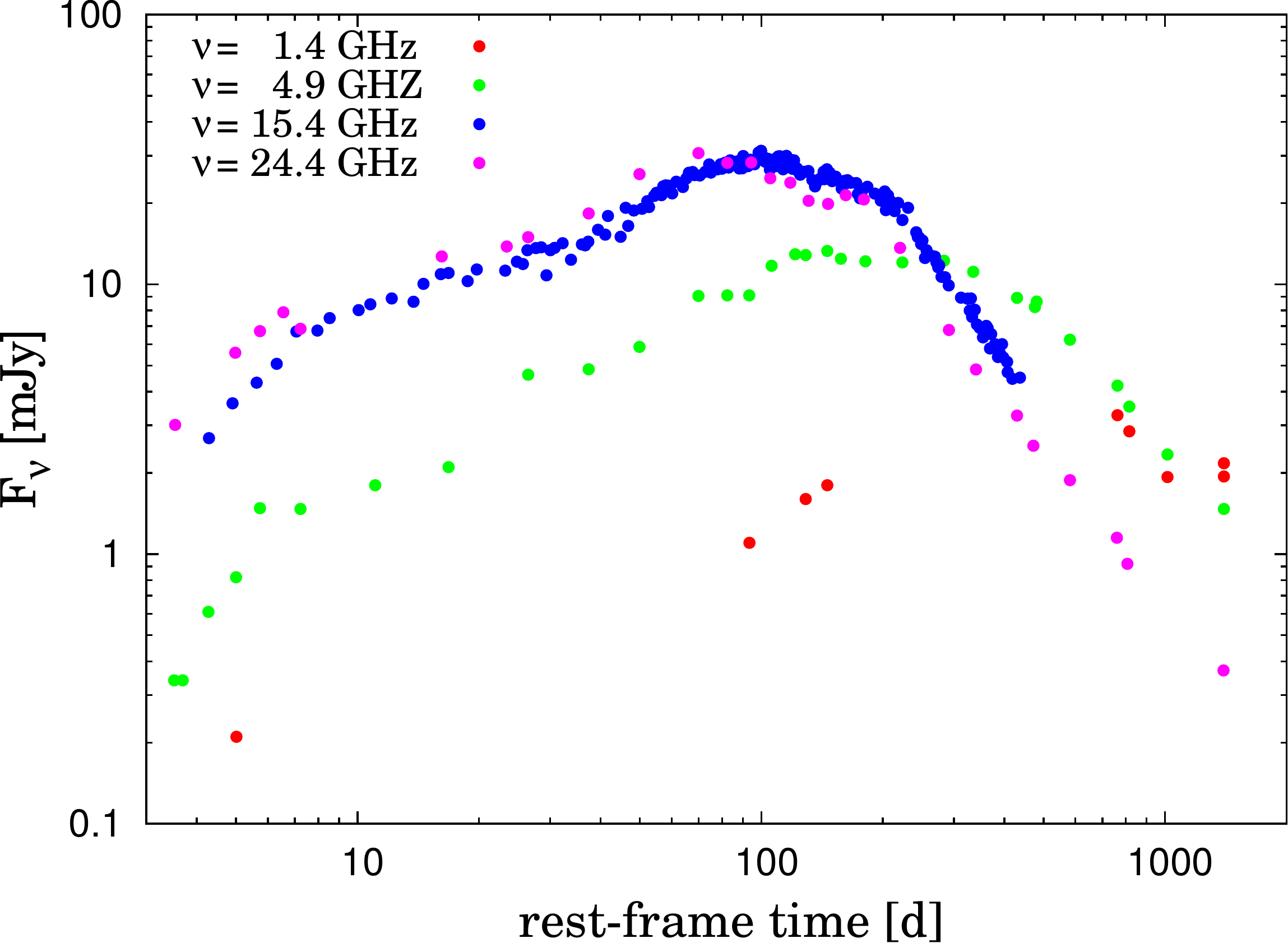}
\caption{Radio lightcurves of the jetted TDE SW1644 at 1.4, 4.9, 15.4 and 24.4 GHz. Observations were done by VLA and are taken from \citet{berger12,zauderer13, eftekhari18}.
}
\label{fig:radio_LC}
\end{figure}

Unlike the fading X-ray emission, the radio flux from SW1644 continued to rise for hundred of days before dropping (see Figure \ref{fig:radio_LC}). The different behaviour with respect to the X-ray emission and the lack of strong variability in radio imply that the two components are likely produced at different locations.

A variety of physical parameters of the system can be inferred from modelling the radio emission, including the shock kinetic energy, shock Lorentz factor, the density stratification of the surrounding medium, and magnetic fields in the shocked region \citep[][]{zauderer11, metzger12, berger12, cao12, barniolduran13, zauderer13, mimica15, eftekhari18}. Although the VLBI Network did not resolve the radio emission region within the first 3 yrs (constraining the transverse size $<2$~pc at $3\sigma$), a statistical astrometric precision of $12\mu$as ($1\sigma$) was obtained, which constrains the apparent proper motion $<0.3c$ at 3$\sigma$ \citep{yang16}. This is consistent with the picture that the observer is near the jet axis (to within a few degrees) and that the blast wave has slowed down to non-relativistic speeds within a few hundred days. In Section 3.2, we provide a simple discussion on modeling the radio emission at late time when the shock has decelerated to non-relativistic speeds.

In addition to SW1644, non-thermal X-ray emission has been detected by \textit{Swift} in other two TDE jet candidates: \textit{Swift} J2058.4+0516 (SW2058 hereafter, \citealt{cenko12}) and \textit{Swift} J1112.2−8238 (SW1112 hereafter, \citealt{brown15, kawamuro16}). These events share many of the characteristics of SW1644: X-ray emission lasting for a month or longer; a power-law decay in the X-ray flux; nearly flat optical-to-X-ray spectral slope $\beta_{\rm OX}\sim 0.1$ at $t\sim 10\,$days; bright radio emission lasting for 3 yrs or longer \citep[late-time radio fluxes for SW2058 and SW1112 are reported by][]{brown17}. SW2058, fainter than SW1644 due to the greater redshift (z = 1.19), showed a decline in the X-ray emission as $\sim t^{-2.2}$ (steeper than that for SW1644) and large variability \citep{cenko12}. The SW1112 X-ray emission, on the other hand, declined as $t^{-1.1}$ with large variability as well \citep{brown17}. As in SW1644, SW2058 (see Figure \ref{fig:X-ray_LC}) showed a sharp drop-off in X-ray flux after a few hundred days \citep{pasham15}, roughly coincident to the expected transition from super-Eddington to sub-Eddington accretion. Due to sparse time sampling, it is unclear whether SW1112 had a rapid X-ray drop on a similar timescale.

\subsection{Other radio jets/outflows}
\label{sec:radio}

In addition to the \textit{Swift} events, four TDE candidates with radio jets/out-flows have been discovered so far:

{\bf ASASSN-14li} is a very close-by ($90$~Mpc) TDE candidate discovered by the \textit{All Sky Automated Survey for Supernovae} \citep{holoien16, brown18}. Radio emission from this source was first detected by \citet{vanvelzen16} at 1.4 GHz with the the \textit{Westerbork Synthesis Radio Telescope} and at 15.7 GHz with the \textit{Arcminute Microkelvin Imager} (AMI), and then in several radio frequencies by \citet{alexander16} and \citet{bright18} using the \textit{Very Large Array} (VLA) and AMI. 

The radio emission shows a characteristic synchrotron self-absorption behaviour at low frequencies ($F_\nu \propto \nu^{5/2}$) and an optically-thin synchrotron emission at high frequencies ($\sim\nu^{-1}$). This behaviour is typically observed in, e.g., radio supernovae and differs from the synchrotron emission from relativistic shocks as the characteristic frequency $\nu_{\rm m}$ (corresponding to the minimum Lorentz factor $\gamma_{\rm m}$ of the shock-accelerated electrons) is below the self-absorption frequency $\nu_{\rm a}$. \citet{alexander16} modeled the observations as due to the forward shock generated by a non-relativistic outflow with kinetic energy of 10$^{47}$--10$^{48}$~erg and velocity $\sim$0.1~c. Under the assumption of equipartition between the energy of non-thermal electrons and magnetic fields \citep{barniolduran13}, they also inferred that the density profile in the environment within $\sim$$0.01$~pc from the SMBH is stratified as $\rho(r)\propto r^{-2.5}$. \citet{alexander16} also pointed out that the unbound debris (about half of the star is ejected at high speeds) typically has a solid angle too small to explain the observations, unless its interaction with the environment forms an extended bow shock with much larger emitting area \citep{krolik16}.

\citet{pasham18} argued that the radio emission is generated internally within an expanding, non-relativistic collimated outflow, based on the discovery of a cross-correlation between the soft X-ray and radio variability with a delay of $\sim$12 days. The size of the synchrotron emitting region (or the ``synchrotron photosphere'') is directly given by the self-absorption frequency $\nu_{\rm a}$ and the flux $F_{\nu_{\rm a}}$ (see discussion in Section 3.2). Taking the delay as the time it takes for the outflow to expand to the size of the ``synchrotron photosphere'', \citet{pasham18} obtained an outflow speed of $\sim$$0.3$--$0.6c$. Observation of an O~VIII absorption feature at blue-shifted velocity of 0.2c supports this picture \citep{kara18}.


The radio emission from ASASSN-14li was resolved by observations by \citet{romero16} employing the \textit{European Very Long Baseline Interferometry Network} (EVN) 200 days after the discovery, showing two spatially separated components connected by faint diffuse emission. The peaks of the two components are separated by a projected distance of 1.9~pc. The brighter, core-like component contributes the majority of the radio flux and is unresolved with size $\lesssim1$~pc. If the fainter elongated component is a jet, then the separation would require superluminal motion at apparent speed 7.4--7.8c and observer's viewing angle less than $\sim$15$^{\rm o}$. This scenario is unlikely because, given the low kinetic energy inferred from the radio flux, the jet should have already decelerated before traveling a physical distance of $\gtrsim7\,$pc. \citet{romero16} prefer the scenario that the elongated component is from a non-relativistic outflow ejected $\gtrsim100$ years ago by historic AGN or TDE activities. An alternative explanation is a binary SMBH.

Despite the debate on the detailed morphology of the radio emission region, the main message from observations of ASASSN-14li is that TDEs can launch non-relativistic outflows. They may come from the tidal unbound debris \citep{krolik16}, the stream self-intersection shocks \citep{2016ApJ...830..125J}, and the accretion disk wind \citep{strubbe09}.

{\bf XMMSL1 J0740-85} is another close-by TDE candidate (75 Mpc) which showed radio spectra consistent with optically thin synchrotron emission \citep{saxton17, alexander17}. Its emission ($\sim$~2--3 yrs after the disruption) is somewhat similar to that from ASASSN-14li and is several orders of magnitude fainter than SW1644 at similar epochs. Due to lack of low-frequency observations, the self-absorption frequency $\nu_{\rm a}$ (corresponding to the peak frequency) was not well measured, which makes it difficult to obtain the total kinetic energy and expansion speed. Adopting an equipartition analysis, \citet{alexander17} constrained the kinetic energy to be in the range $5\times10^{49}$--$4\times 10^{51}\,$erg and the ambient medium density to be in the range $0.03$--$7\times10^{4}\rm\,cm^{-3}$ (assuming a constant density medium).

{\bf Arp 299-B AT1}. \citet{mattila18} presented observations of a spatially resolved jet structure associated with a bright infrared flare in the nucleus of the Arp 299 galaxy (at 44.8 Mpc), which is a dusty luminous infrared galaxy (LIRG) undergoing a major merger. The flare radiated $1.5\times 10^{52}$~erg of energy in the near-infrared (even without bolometric correction), and the SED is consistent with the thermal emission from circum-nuclear dust heated to $T\sim10^3\,$K by the UV-optical emission from a TDE \citep{lu16, 2016ApJ...828L..14J, 2016ApJ...829...19V, 2018MNRAS.477.2943W}. Due to the large dust column ($A_{\rm V}\sim 460$~mag), the source was not detectable in the optical or soft X-ray bands. The radio flux at 8.5~GHz increased in the first few hundred days and then steadily declined in the subsequent observations over $10$~yrs. VLBI observations showed that the source was initially compact and then gradually developed a one-sided jet structure moving at an averaged apparent transverse speed of $0.25 c$ \citep{mattila18}. Non-detection of a counter-jet constrains the observer viewing angle to be $\sim$ 30$^{\rm o}$ from the jet axis, provided that the emission from the counter-jet is not absorbed by the dense circum-nuclear medium. \citet{mattila18} also provided a hydrodynamic modeling of the apparent motion and the radio SED and found that the data can be fitted with a relativistic jet with kinetic energy of $\sim$$10^{51}$~erg.

{\bf IGR J12580+0134} was discovered as a hard X-ray flare by \textit{INTEGRAL} at the center of the close-by (17 Mpc) Seyfert 2 galaxy NGC 4845 \citep{nikolajuk13}. Rapid rise in the soft X-ray flux and then a gradual decay were detected by \textit{XMM-Newton}, \textit{Swift} and \textit{MAXI} \citep{nikolajuk13, kawamuro16}, with total radiated energy of $\sim$$10^{49}$~erg 
(or $10^{50}$~erg for a bolometric correction factor of 10). 
Radio \citep{irwin15, lei16} and millimetre emission \citep{yuan16} were detected and interpreted by these authors as generated by an off-axis relativistic jet
(ejected as the result of  the tidal disruption of a star with a mass of 8-40 Jupiter)
interacting with the surrounding medium. However, as discussed by \citet{auchettl17}, persistent radio emission was detected from this source before the X-ray flare and showed large amplitude variability historically indicating that this source presents AGN activity. Late-time VLBI observations show an extended disk-like structure with a projected length of $\sim$ $3$~pc, which may be related to nuclear star formation \citep{perlman17}. Another resolved source at a projected distance of 4.1~pc is likely unrelated to this flare \citep{perlman17}.


\subsection{What fraction of TDEs have relativistic jets?}
\label{sec:rate}

In this subsection, we conservatively estimate the jet fraction for TDEs $f_{\rm j}$, with the goal of understanding what parameters control the on-and-off switch for jet launching. Our conclusion is that this fraction is so-far only weakly constrained to be $3\times10^{-3} \lesssim f_{\rm j} <1$.

\textit{Swift} detected only three events in $\sim$$10 $~yrs, which implies\footnote{The peculiar fact that all three events were discovered in a period of three months in 2011 seems to be due to a statistical fluke \citep{brown17}.} a rate $\sim 3\times 10^{-11}\rm \,Mpc^{-3} \,yr^{-1}$ \citep{burrows11, brown15}. In Section 3.2, we show that the $\gamma$-ray beaming factor for SW1644 is of order $f_{\rm b}\sim 0.01$. Taking a total TDE rate of order $10^{-6}\rm\,Mpc^{-3}\, yr^{-1}$ \citep{2018ApJ...852...72V}, we obtain the jet fraction $f_{\rm j}\sim 3\times10^{-3} (f_{\rm b}/0.01)^{-1}$. We also note the selection bias of \textit{Swift} against relatively faint, long transients such as ultra-long GRBs and jetted TDEs \citep[they are less likely to trigger BAT, see][]{levan14}. For instance, the discovery of SW2058 and SW1112 were made possible by four-day integration of the $\gamma$-ray flux \citep{cenko12, brown15}. Jetted TDEs with a slightly softer high-energy spectrum or shorted peak duration may fail to reach the significance for triggering and hence have been missed. Currently, it is difficult to quantify the incompleteness correction without a larger sample of SW1644-like events. Nevertheless, this correction factor will increase the jet fraction of TDEs, and hence $f_{\rm j}\gtrsim 3\times10^{-3}$.

For a relativistic source, the flux for an off-axis observer is suppressed by a factor of $(\Gamma\theta_{\rm obs})^{-6}$ compared to the on-axis flux, where $\Gamma$ is the Lorentz factor of the source and $\theta_{\rm obs}$ is the observer's viewing angle \citep[e.g.][]{granot02}. Off-axis jets from TDEs, which should dominate the population of existing events, are then much fainter in hard X-rays/$\gamma$-rays.

On the other hand, when the jet decelerates to non-relativistic speeds on a timescale of $\sim$$1$~yr, the radio emission from the forward shock should be roughly isotropic. Thus, the radio flux from the off-axis cousins of SW1644 should be at the level of 1--10~mJy a few years after the discovery \citep{giannios11}. Several authors have looked for radio emission from TDEs \citep[e.g.,][]{grupe99, komossa02, saxton12, bower13, vanvelzen13, chornock14, blagorodnova17, mattila18} and only in one or two cases \citep{bower13, mattila18} there is evidence of off-axis relativistic jets. At face value, this implies that only a few percent TDEs produce relativistic jets. However, the flux at a given radio frequency for an off-axis observer depends sensitively on the unknown circum-nuclear medium density profile, magnetic field strength, and the angular structure of the jet. The surveys by \citet{bower13} and \citet{vanvelzen13} typically only had single-epoch upper limits at one or two frequencies, so the existence of relativistic jets, especially somewhat weaker ones than that in SW1644, cannot be confidently ruled out. For instance, \citet{generozov17} explored a wide range of circum-nuclear densities and used these radio upper limits to constrain the kinetic energy of possible relativistic jets to be $E_{\rm k}\lesssim 10^{52}$--$10^{53}$~erg.

The constraints are stronger for very nearby TDEs with multi-epoch follow-up observations. For instance, the radio afterglow from ASASSN-14li suggests a non-relativistic outflow of kinetic energy of order $\sim$$10^{48}\,$erg.
XMMSL1 \citep[][]{alexander17} showed similar properties as ASASSN-14li and iPTF16fnl \citep{blagorodnova17} had slightly more stringent radio upper limits. Thus, we conclude that the jet fraction for TDEs is larger than $3\times10^{-3}$ but not universal, which implies that super-Eddington accretion\footnote{The peak accretion rate during a TDE is typically larger (by several orders of magnitude) than the Eddington limit (e.g., \citealt{guillochon13, lu18}).} is not a sufficient condition for launching relativistic jets.

We conclude this subsection by noting that TDEs seem to launch two different categories of outflows: collimated relativistic jets with Lorentz factor $\Gamma \sim$ a few--$10$ (the \textit{Swift} events and perhaps Arp 299-B AT1); and non-relativistic outflows (ASASSN-14li and possibly many other TDEs). This dichotomy is somehow similar to that between SNe and GRBs (see, e.g., figure 6 of \citealt{alexander16}), where a large fraction of events produce non-relativistic flows while relativistic jets are rare. Currently, it is unclear what mechanisms regulate the generation of relativistic jets vs. non-relativistic outflows. Extensive radio follow-up studies of a sample of nearby TDEs should provide an answer to this question.

\section{Models}
\label{sec:models}

In this section we will briefly review the dynamics of TDE jets (\S 3.1) and the radiative processes responsible for the observed non-thermal radio (\S 3.2) and X-rays (\S 3.3) emission.

\subsection{Jet dynamics}

It has been proposed that efficient jet launching can be achieved in the presence of a large magnetic flux threading a rapidly spinning BH \citep[e.g.,][]{blandford77, tchekhovskoy11}. In terms of TDE jets, \citet{tchekhovskoy14} showed that a strong poloidal magnetic field is needed to explain observations of the SW1644 event, and that the large flares observed in X-rays in SW1644 are generated by the jet wobbling violently before rearranging itself into a ``magnetically arrested disk'' state. \citet{curd19} presented general relativistic radiation MHD simulations of accretion onto a BH during a TDE, showing that a strong poloidal magnetic field and rapid BH spin are needed to reproduce the jet power in SW1644. It is unclear whether large poloidal magnetic fields can be generated by the turbulent dynamo inside the accretion flow or it requires preexisting net poloidal magnetic flux in the infalling matter \citep[e.g. the disrupted star,][]{guillochon17, bonnerot17}. \citet{kelley14} showed that the magnetic flux in a fossil disk (due to pre-TDE AGN activity) can be dragged to near the BH by the fallback stream. More recent high-resolution (comparable to typical local shearing-box simulations) global general relativistic MHD simulations show that turbulent dynamo can indeed generate poloidal magnetic flux in situ, even for the initial condition of purely toroidal magnetic fields \citep{liska18}.

The dynamics of TDE jet propagation on large scales has been studied both analytically \citep[e.g.,][]{giannios11, metzger12} and numerically \citep{decolle12,mimica15, generozov17} and is briefly reviewed in this sub-section. As the plasma ejected from the ``central engine'' (the BH-disk system) interacts with the ambient medium, it forms a double-shock structure (the ``working surface'' or the ``jet head") made by the forward shock, which heats and accelerates the ambient medium, and the reverse shock, which is heating and decelerating the jet material.

We denote quantities in the unshocked fluid with subscript $_1$  and those in the shocked fluid with subscript $_2$ (each in their comoving frame). The Rankine-Hugoniot jump conditions give the post-shock density $\rho_2= 4\Gamma_{\rm rel} \rho_1$ and pressure $p_2= (4/3)(\Gamma_{\rm rel}^2-1) \rho_1 c^2$, where $\Gamma_{\rm rel}$ is the relative Lorentz factor between the post-shock and pre-shock fluid and we have assumed the pre-shock fluid to be cold (or pressureless). The double-shock structure is connected by by a contact discontinuity which separates the shocked jet material and the shocked ambient medium. The pressure and velocity are continuous across the contact discontinuity. If the inertia of the jet head is negligible, i.e. if changes in parameters near the forward shock produce instantaneous changes near the reverse shock, then the two shocked regions have the same pressure and velocity \citep[e.g.,][]{sari95, uhm11}. The relative Lorentz factor between the jet head and the unshocked jet is $\gamma_{\rm rel} = \Gamma_{\rm j}\Gamma_{\rm h} (1 - \beta_{\rm j}\beta_{\rm h})$ and the pressure of the jet head is $p_{\rm h} = (4/3)(\gamma_{\rm rel}^2-1)\rho_{\rm j} c^2$, where $\Gamma_{\rm j}$ is the jet Lorentz factor and $\rho_{\rm j}$ is the jet density. On the other hand, the pressure of the jet head is also given by the jump conditions at the forward shock $p_{\rm h} = (4/3)(\Gamma_{\rm h}^2-1)\rho_{\rm a}c^2$. Note that pressures and densities are measured in the comoving frame of the local fluid. Thus, the velocity of the jet head can be solved to be
\begin{equation}
\label{eq:vhead}
v_{\rm h} = \frac{v_{\rm j}}{1+
\left(\rho_{\rm a}/\rho_{\rm j} \Gamma_{\rm j}^2 \right)^{1/2}}.
\end{equation}

Taking the density stratification of the ambient medium to be $\rho_{\rm a} = \rho_0 (r/r_0)^{-k}$ and the jet density to be $\rho_{\rm j}= L_{\rm j}/(4\pi r^2 c^3\beta_{\rm j}\Gamma_{\rm j}^2)$ ($L_{\rm j}$ being the isotropic equivalent power for a conical jet), one can show that the jet head velocity will increase/drop in regions in which the ambient medium drops faster/slower than $r^2$. Eq. \ref{eq:vhead} shows that the dynamics of the jet head is mainly regulated by the jet kinetic luminosity $L_{\rm j}$ and the ambient density $\rho_{\rm a}$\footnote{Other jet parameters (e.g., the angular energy distribution, the magnetic fields) have not been considered in detail in the literature in the context of TDE jets. In particular, pinch/kink instabilities due to the presence of dynamically important magnetic fields could dramatically change this simple picture.}. In the following, we discuss how the dynamics of the jet propagation is regulated by the injection history and density stratification of the ambient medium.

Without a physical model of the jet launching process, it is difficult to know the time-dependent jet luminosity from first principle. A simple prescription is to take the jet luminosity to be a constant fraction (e.g. 10\%) of the well-known mass fallback rate \citep{decolle12}. Alternatively, if the jet converts a constant fraction of its kinetic energy into X-rays as seen in SW1644, then one can infer that the jet power was roughly constant for a duration $t_{\rm j}\sim 10$~days and then dropped as $t^{-5/3}$. However, it is likely that the \textit{Swift} trigger was substantially delayed (by e.g. several weeks) with respect to the initial jet launching \citep{decolle12, tchekhovskoy14}. 


The ambient medium structure within $\sim$$10^{15}\,$cm from the BH is shaped by the TDE debris. The main processes are stream self-intersection, secondary shocks, and viscous accretion \citep{2015ApJ...804...85S, 2016ApJ...830..125J, 2016MNRAS.458.4250S}, but it is currently too computationally expensive to perform global numerical simulations of TDEs. \citet{loeb97} described a simple constant entropy envelope supported by radiation pressure gradient against gravity, with density profile given by $\rho_{\rm a} = M_{\rm env}/[4\pi r^3 \mr{ln}(r_{\rm sca}/r_{\rm t})]$. The inner boundary is given by the tidal radius $r_{\rm t}$ and the outer boundary is the electron scattering photosphere $r_{\rm sca}\sim 1.5\times10^{15} \mr{\,cm}\, (M_{\rm env}/0.5\msun)^{1/2}$. The jet propagating through such an envelope is strongly decelerated in that $v_{\rm h}\ll v_{\rm j}$, and the jet would cross this region in $t_{\rm cross}\sim1\mr{\,day}\, (M_{\rm env}/0.5\msun)^{1/2}(L_{\rm j}/10^{48}\rm\,erg\,s^{-1})^{-1/2}$. Numerical simulations by \citet{decolle12} show that the jet crosses the envelope formed around a 10$^7$ M$_\odot$ in $\sim$12 hrs.

Assuming an infinitely thin working surface, the rarefaction wave arrives at the forward shock at a time $t=v_{\rm j} t_{\rm j}/(v_{\rm j} - v_{\rm h}) \approx (1+\beta_{\rm h})\Gamma_{\rm h}^2 t_{\rm j}\simeq t_{\rm j}$. Thus, a jet may be hydrodynamically choked if the injection duration $t_{\rm j}$ is shorter than the crossing time $t_{\rm cross}$, so we obtain the choking condition to be $L_{\rm j}\lesssim 10^{46}\mr{\,erg\,s^{-1}}\, (M_{\rm env}/0.5\msun) (t_{\rm j}/10\mr{\,day})^{-2}$ \citep[see also][]{wang16}. Realistically, the envelope mass may be significantly less than $0.5\msun$ when the jet launching starts. Late-time jet injection as seen in SW1644 $L_{\rm j}\propto t^{-5/3}$ makes it (slightly) easier for the jet the punch through. Therefore, we see that only the weakest jets may be hydrodynamically choked as it may occur in e.g., GRBs. The majority of the relativistic jets in TDEs should be able to propagate out to much larger distances, and the energy deposed by the jet into a cocoon ($E_{c}\sim 10^{50}-10^{51}$~erg) should quickly unbind the envelope.

\begin{figure}
\centering
\includegraphics[scale=0.9]{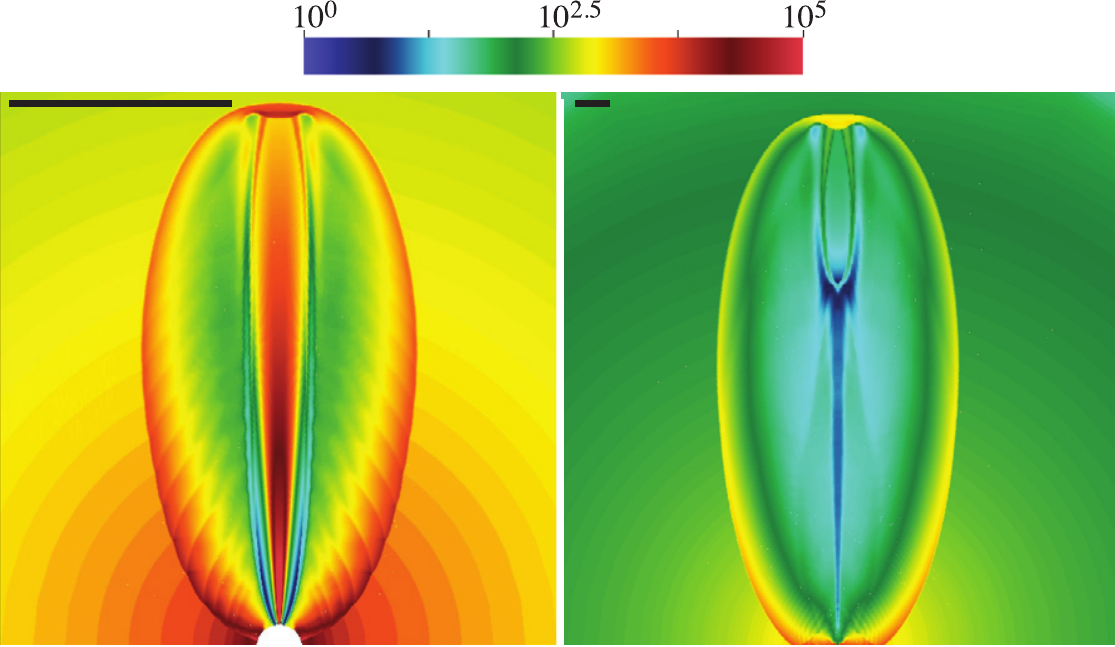}
\caption{Density stratification (in units of cm$^{-3}$) of a relativistic jet propagating through the region shaped by the interaction of stellar winds around a SMBH (Figure taken from \citealt{decolle12}). On the right panel, the jet channel is partially filled once the jet is switched-off after the  accretion becomes sub-Eddington. At later times, the rarefaction wave will arrive at the working surface and the reverse shock will disappear. When the jet starts to decelerate significantly, the structure will then converge to the \citet{blandford76} solution. Each panel shows a $10^{17}$ cm scale bar on the upper left corner.
}
\label{fig:sim}
\end{figure}

Once the jets break out from the inner region shaped by the tidal debris, it will accelerate to relativistic speed and propagate through the environment of the SMBH shaped by the winds of massive stars (see Figure \ref{fig:sim}). The density distribution goes as $\rho\propto r^{-1.1}$ to $r^{-1.5}$ depending on the SMBH mass and the energy injection in the winds \citep{quataert04, decolle12,generozov17}. In the outer regions ($r \gtrsim 1$~pc) the density converges to the $r^{-2}$ wind profile. Once the rarefaction wave arrives to the head jet, the impulsive jet/shell undergoes a coasting phase with constant velocity until enough mass $\sim E_{\rm j} /\Gamma_{\rm j}^2 c^2$ has been swept up ($E_{\rm j} = \int L_{\rm j}\d t$ being the isotropic equivalent energy). Then, the hydrodynamic structure approaches the \citet{blandford76} self-similar solution. When the shock decelerates to a mildly relativistic Lorentz factor $\Gamma\sim 2$, lateral expansion becomes important and further deceleration to non-relativistic speeds occurs rapidly \citep{2012MNRAS.421..570G}. At very late time, greater than ($E_{\rm k}$ being the total kinetic energy of the jet)
\begin{equation}
 t_{\rm NR} \sim 240
 \left(
 \frac{E_{\rm k}}{10^{53}\, \rm erg}
\right)^{1/3} 
   \left(
 \frac{n}{100\, \rm cm^{-3}}
 \right)^{-1/3} 
 \; {\rm days}.
\end{equation}
the jet will finally decelerate to non-relativistic speeds and spread laterally into a roughly spherical shell.



\subsection{The Radio Afterglow of SW1644}
\begin{figure}[htbp!]
\centering
\includegraphics[scale=0.3]{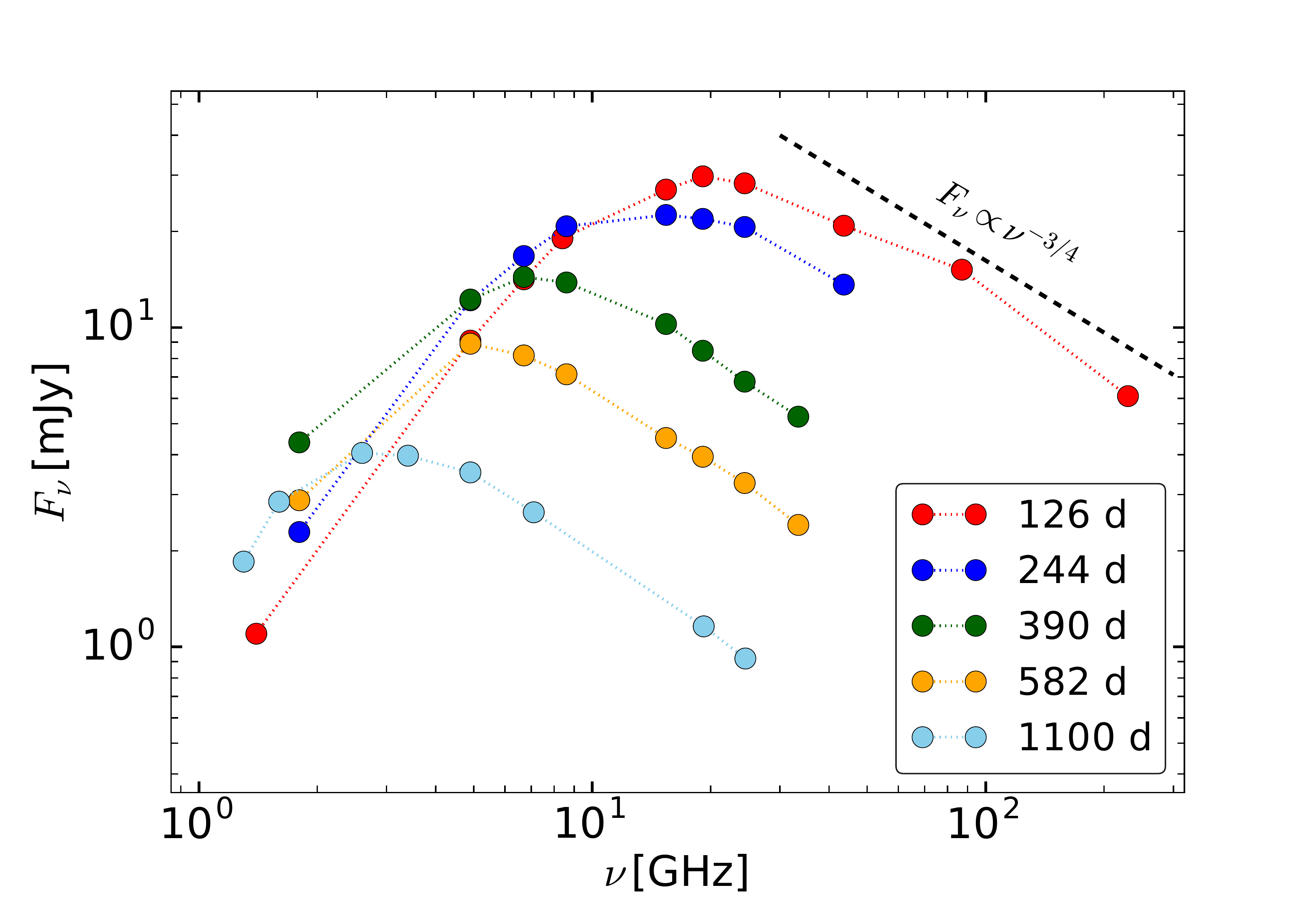}
\caption{Late-time radio SEDs for SW1644 from $\sim$100 to 1000 days, taken from \citet{eftekhari18}. The black dashed line shows the high-frequency synchrotron spectrum for a power-law distribution of electrons $F_\nu\propto \nu^{(1-p)/2}$, and an electron power-law index of $p\approx 2.5$ reasonably agrees with the data.}
\label{fig:radio-SED}
\end{figure}

\begin{figure}[htbp!]
\centering
\includegraphics[scale=0.25]{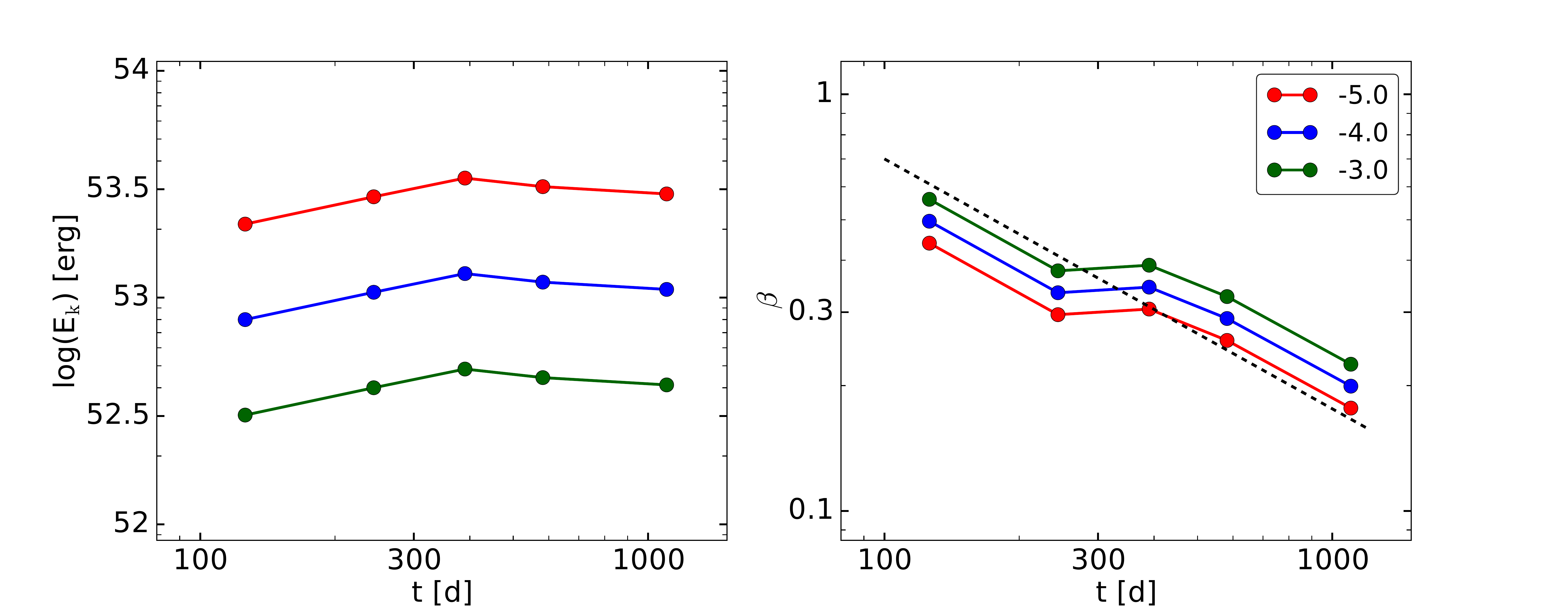}
\caption{\textit{Left Panel}: Radio calorimetry measuring the total kinetic energy of the ejecta $E_{\rm k}$ from SW1644. Different colors are for different magnetic energy fractions $\epsB = 10^{-5}$ (red), $10^{-4}$ (blue), $10^{-3}$ (green). We keep the electron energy fraction $\epse=0.1$ fixed. \textit{Right Panel}: Expansion speed of the shocked region $\beta$ varying with time. The black dashed line shows a power-law of $\beta\propto t^{-3/5}$, as expected for a Sedov-Taylor blast wave propagating in a uniform medium.}
\label{fig:shock-energy}
\end{figure}

In this subsection, we do not attempt to include the full hydrodynamics of the 
jet\footnote{Numerical simulations by \citealt{mimica15} showed that  a ``structured'' jet, with an inner core moving with a Lorentz factor of $\sim 10$ and an outer sheath moving with a Lorentz factor of $\sim 2$ is consistent with radio observations of SW1644.} and possibly other ejecta components interacting with the spatially non-uniform circum-nuclear medium (CNM). The early-time afterglow may be affected by many processes, including relativistic beaming, multiple-zone emission, jet angular structure \citep{mimica15}, external inverse-Compton cooling \citep{kumar13}, etc. Our goal is to estimate the total kinetic energy $E_{\rm k}$ injected into the CNM by looking at the late-time radio data (see Figure \ref{fig:radio-SED}). We follow the standard radio calorimetry method and the result is of order of $10^{53}$~erg, as shown in Figure  \ref{fig:shock-energy}. Our calculations are carried out in the host-galaxy rest frame.

The radio SEDs of SW1644 at all epochs clearly show self-absorption below the characteristic frequency $\nu_a$. Analogous to inferring the stellar radius by measuring the luminosity and temperature of a star, the flux $F_{\nu_a}$ tells us the surface area of ``synchrotron photosphere'' which is related to the expansion history of the forward shock. At sufficiently late time ($t\gtrsim 100\,$days), when the ejecta has decelerated to non-relativistic speeds, the forward shock relaxes to the spherical Sedov Taylor self-similar solution. When the shock is at radius $r$ from the center of explosion, the specific luminosity at the self-absorption frequency $\nu_a$ is given by
\begin{equation}
    \label{eq:Lnua}
    \nu_a L_{\nu_a} \simeq 4\pi^2 r^2 \frac{2\nu_a^3 kT}{c^2},
\end{equation}
where $kT\simeq \ga \me c^2$ is the temperature of electrons responsible for absorption, $\ga$ be the Lorentz factor, and $\me$ is the electron mass. At time $t$, the emitting radius $r$ is related to the expansion speed $\beta c$ by
\begin{equation}
    \label{eq:radius}
    r \simeq \beta c t.
\end{equation}
The electron number density in the shocked region is denoted as $n_{\rm sh}$ (which is a factor of 4 higher than that in the pre-shock region), and the magnetic field strength can be written as
\begin{equation}
    \label{eq:Bfield}
    B = (4\pi \epsB \beta^2 n_{\rm sh} \mp c^2)^{1/2},
\end{equation}
where $\epsB\ll 1$ is the fraction of thermal energy density ($e = \beta^2 n m_{\rm p} c^2/2$) taken by magnetic fields, and $\mp$ is proton mass. The electron Lorentz factor corresponding to synchrotron frequency $\nu_a$ is given by
\begin{equation}
    \label{eq:ga}
    \ga = \left(\frac{4\pi \me c \nu_a }{3 eB}\right)^{1/2},
\end{equation}
where $e$ is electron charge.
We assume electrons are accelerated by the shock into a relativistic power-law distribution $\d n/\d \gamma = (p-1) f_{\rm rel} n_{\rm sh} \gm^{p-1} \gamma^{-p}$ for $\gamma>\gm$, and the minimum Lorentz factor $\gm$ and the fraction of electrons accelerated to relativistic speeds $f_{\rm rel}$ are given by
\begin{equation}
    \label{eq:gm}
    \gm = \mr{max}\left[1, {\beta^2\over 2} {p-2\over p-1}{\epse \mp \over \me}\right], f_{\rm rel} = \mr{min}\left[1, {\beta^2\over 2} {p-2\over p-1}{\epse \mp \over \me} \right],
\end{equation}
where $\epse\ll 1$ is the fraction of thermal energy taken by the accelerated electrons. Note that we have accounted for the deep Newtonian case where only a fraction $f_{\rm rel}<1$ of the shocked electrons have relativistic energies. The characteristic synchrotron frequency of electrons at $\gm$ is denoted as $\nu_m$. The late-time radio SED indicates $\nu_a > \nu_m$ and hence $\ga > \gm$. The radiation intensity at $\nu_a$ is related to the number of electrons with Lorentz factor near $\ga$ in the following way
\begin{equation}
    \label{eq:intensity}
    {2\nu_a^2 kT \over c^2} \simeq 2\nu_a^2 \ga \me \simeq {1\over 4\pi} {e^3B\over \me c^2} \left.\left(\gamma {\d n'\over \d\gamma}\right)\right|_{\ga} \Delta \ell_r,
\end{equation}
where $\Delta \ell_r \simeq r/10$ is the radial thickness of the emitting region.

Therefore, we have six equations (\ref{eq:Lnua}--\ref{eq:intensity}) for six unknowns $\beta, r, n_{\rm sh}, B, \ga, \gm$, provided the shock microphysics parameters $\epse$ and $\epsB$ are taken from other independent studies \citep[of e.g. GRB afterglows][]{kumar15}.
The electron power-law index is directly obtained to be $p\approx 2.5$ from the high-frequency radio SED which goes as $F_\nu\propto \nu^{(1-p)/2}$ \citep[in the slow cooling regime][]{2002ApJ...568..820G}. The observables are expressed conveniently in the following way: $\nu_a L_{\nu_a} = 10^{42}L_{a,42}\mr{\,erg\,s^{-1}}$, $t = 300 t_{300}/(1+z)\,$days, $\nu_a = 10\nu_{a,10}(1+z)\,$GHz. Taking $z=0.354$ for SW1644, the final solution is
\begin{equation}
\begin{split}
    \beta &\simeq 0.39\, \epsilon_{e,-1}^{-0.08}\epsilon_{B, -3}^{0.05} {L_{a,42}^{0.45} \over t_{300}^{0.95} \nu_{a,10}^{1.40}},\\
    r & \simeq (5.7\times10^{17}\mr{\,cm})\, \beta t_{300},\\
    n_{\rm sh} &\simeq (2.9\times10^4\mr{\,cm^{-3}}) \,\epsilon_{e,-1}^{-0.47} \epsilon_{B, -3}^{-0.68} {t_{300}^{2.3} \nu_{a,10}^{5.6} \over L_{a,42}^{1.32}},\\
    B &\simeq (4.3\times10^{-3}\mr{\,G})\, \epsilon_{B, -3}^{1/2}\beta n_{\rm sh}^{1/2},\\
    \ga &\simeq 17.0 {L_{a,42} \over \beta^2 t_{300}^2 \nu_{a,10}^2},\\
    \gm &\simeq 30.6\, \epsilon_{e,-1} \beta^2.
\end{split}
\end{equation}
We note that the above formalism only applies to non-relativistic shocks where all electrons are accelerated to relativistic energies, i.e. $\beta^2 \ll 1$ and $\gm \gtrsim 2$. Therefore, we should only use the radio data for intermediate timescales $\sim$100--1000 days, taken from \citet{eftekhari18}. Instead of carrying out SED fitting, we simply take the peak frequency to be $\nu_a$ and the peak flux density to be $F_{\nu_a}$ at each epoch, affording an order unity error. After solving all the variables, the total kinetic plus thermal energy in the shocked region is given by
\begin{equation}
    E_{\rm k}\simeq 4\pi r^2 \Delta\ell_r {\beta^2\over 2} n_{\rm sh} \mp c^2 \simeq (5.0\times10^{52}\mr{\,erg})\, \epsilon_{e,-1}^{-0.87}\epsilon_{B, -3}^{-0.42} 
    {t_{300}^{0.58} L_{a,42}^{0.92} \over \nu_{a,10}^{1.34}}.
\end{equation}
The electron energy fraction $\epse\sim 0.1$ is fairly well established, but the uncertainty on the magnetic energy fraction $\epsB$ may be up to a few orders of magnitude. Thus, we consider three different values $\epsB=10^{-5},\, 10^{-4},\, 10^{-3}$ in Figure \ref{fig:shock-energy}, and the total kinetic energy is constrained to be $\sim$ $10^{53}\,$erg, in agreement with \citet{barniolduran13}.

\subsection{The X-ray Emission Mechanism}

In this subsection, we discuss the possible mechanisms for generating the extremely bright X-ray emission from jetted TDEs. We focus on SW1644 due to its rich set of data (or flux limits) from radio to TeV bands. For convenience, frequency and time are measured in the host-galaxy rest frame.

The X-ray emission has averaged isotropic luminosity $L_{\rm X}\sim 10^{47} \rm\,erg\,s^{-1}$ in the first 10 days and then declines roughly as $t^{-5/3}$ until a sudden drop into a plateau at about $t=370$ days (in the host-galaxy frame). The X-ray spectrum in the $0.3(1+z)${--}$10(1+z)\,$keV range is a power-law $F_\nu\propto \nu^{-\alpha}$ with spectral index $\alpha\simeq 0.8$ at early time (1{--}2 months), and then it hardens to $\alpha\simeq 0.5$ at $\sim$100 days. The early-time spectrum extends to at least $50(1+z)\,$keV without a break \citep{levan11, burrows11}, which means most radiation energy is released at even higher frequencies.

In the high-energy gamma-ray band ($0.1(1+z)$--$10(1+z)\,$GeV), \textit{Fermi} LAT provides stringent constraints on the jet gamma-ray luminosity\footnote{Similar constraints have been put on SW2058 and SW1112 as well \citep{peng16}.}, i.e. $L_{\rm LAT}<1\times10^{46}\rm\,erg\,s^{-1}$ within the first day, $<3\times10^{45}\rm\,erg\,s^{-1}$ within the first 8.5 days, and $<5\times10^{44}\rm\,erg\,s^{-1}$ within the first 85 days \citep{peng16}. In the very-high-energy gamma-ray band (0.1{--}$1\,$TeV), \textit{MAGIC} provided constraints $\nu L_\nu (0.14(1+z)\,\mr{TeV})<5\times10^{45}\rm\,erg\,s^{-1}$, $\nu L_\nu (0.32(1+z)\,\mr{TeV})<1.6\times10^{45}\rm\,erg\,s^{-1}$, $\nu L_\nu (0.65(1+z)\,\mr{TeV}) <4\times10^{44}\rm\,erg\,s^{-1}$ within the first 13 days \citep{aleksic13}. The limits from \textit{VERITAS} \citep{aliu11} are similar to those of \textit{MAGIC}.

\begin{figure}
\centering
\includegraphics[scale=0.27]{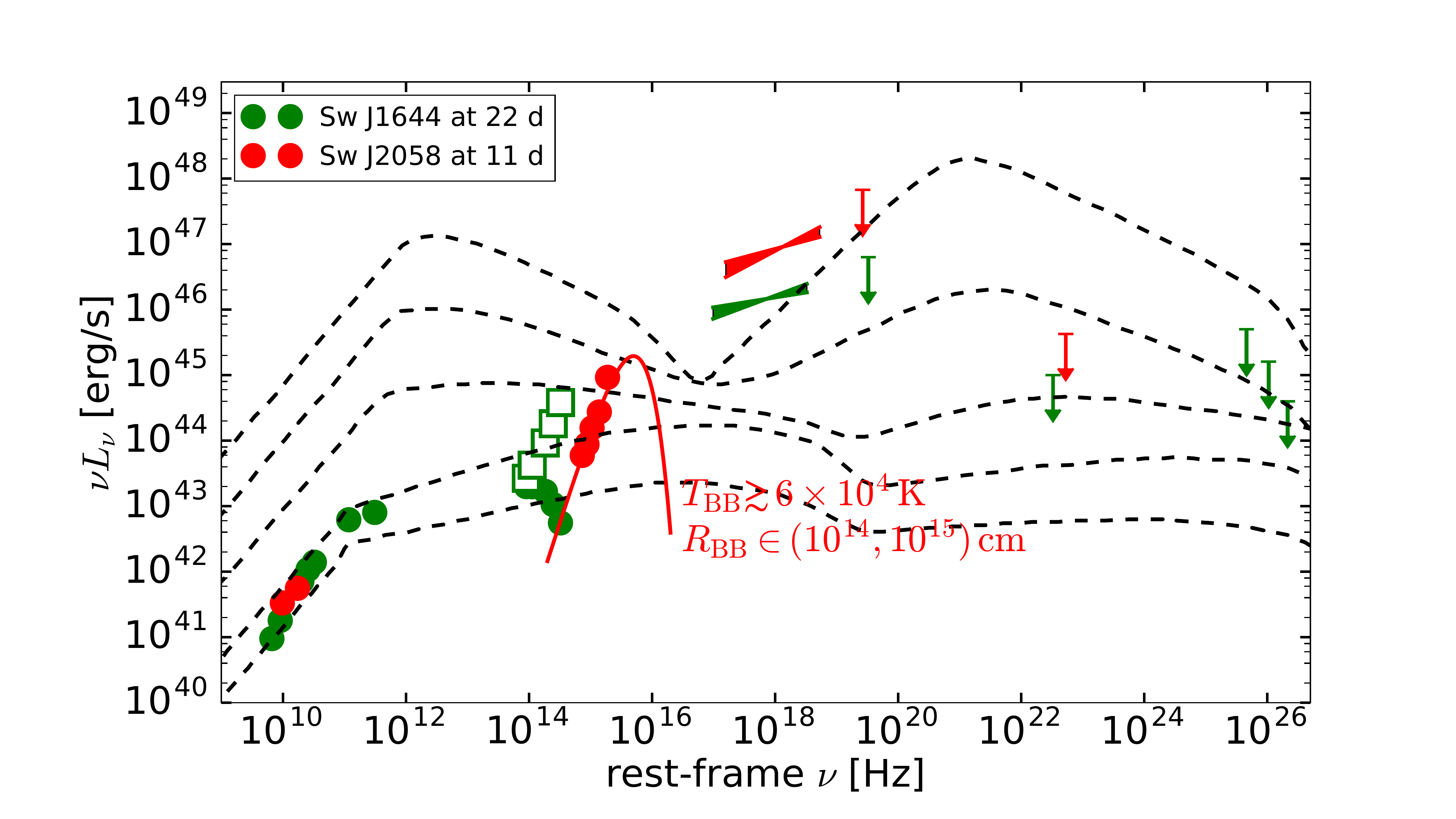}
\caption{Comparison between the SEDs of jetted TDEs with the blazar sequence from \citet{ghisellini17}, which considers a larger sample with more complete multi-wavelength data than the original version by \citet{fossati98}. The \textit{Swift} BAT gamma-ray flux at $50(1+z)\,$keV for SW1644 has large uncertainties due to low photon statistics \citep{burrows11, levan11}. The near-infrared flux at $\sim10^{14}\,$Hz for SW1644 suffers large dust extinction, and the green squares show the corrected fluxes assuming $A_{\rm V}=4.5\,$mag. The most important two differences are the lack of emission in the near-infrared band ($\sim10^{14}\,$Hz) and high energies $\gtrsim0.1\,$GeV ($\gtrsim3\times 10^{22}\,$Hz) in jetted TDEs.}
\label{fig:SED}
\end{figure}

In figure \ref{fig:SED} we compare the broad-band SEDs of SW1644 and SW2058 in the first $\sim$ 10 days with the blazar sequence by \citet{ghisellini17}. Two important differences can be noticed: (1) SW1644 and SW2058 have exceptionally low ratios $L(>\,0.1\mr{\,GeV})/L_X \lesssim$ a few percent, while for blazars this ratio is typically of order unity. (2) The strong infrared (IR) emission typically seen from blazars is missing in SW1644 and SW2058. For SW1644, the K-band (rest-frame frequency $1.8\times10^{14}\,$Hz) luminosity is $\nu L_{\nu}(\mbox{K-band})\simeq 2\times10^{43}\rm \,erg\,s^{-1}$ between 3 and 30 days \citep{levan16}. Since dust extinction is weak at K-band, the IR-to-X-ray luminosity ratio is $L(\mbox{K-band})/L_X \lesssim 10^{-4}$. These two differences are interesting and mysterious. They may provide some clue on the composition and radiation mechanism for relativistic astrophysical jets in general, if TDE jets are representative of the entire family.

In the following analysis, we take a quasi-steady jet with bulk Lorentz factor $\Gamma= 10 \Gamma_1$ and isotropic luminosity $L_{\rm j} = 10^{48}L_{\rm j,48}\rm\,erg\,s^{-1}$. The opening angle of the jet is assumed to be close to or larger than $\Gamma^{-1}$, so the boundary in the transverse direction is not important (due to de-beaming). We assume a fraction $\xi_{\rm B}$ of the total jet luminosity is in the form of Poynting power, $L_{\rm B} = \xi_{\rm B}L_{\rm j}$. The jet magnetization parameter $\sigma$ is defined as the ratio between the Poynting flux and particles' kinetic flux carried by the jet, so we have $\xi_{\rm B} = \sigma/(1+\sigma)$. The X-ray flux and spectrum for SW1644 were highly variable and strongly affected by neutral hydrogen absorption. We take a single power-law $L_\nu\propto \nu^{-0.8}$ representative of the early-time ($\lesssim 10\,$d) overall behavior \citep{saxton12}. The total isotropic luminosity in the $0.3(1+z)$--$10(1+z)\,$keV band is denoted as $L_X\equiv \int_{0.3(1+z)}^{10(1+z)}  L_\nu\d \nu = 10^{47}L_{X,47}\rm\,erg\,s^{-1}$, and the inverse relation is $\nu L_\nu = 2.36\times10^{46} \mr{\,erg\,s^{-1}} \,\nu_{\rm keV}^{0.2} L_{X,47}$.

In the following, we consider three popular radiation mechanisms for the X-ray emission: synchrotron, synchrotron self-Compton (SSC), and external inverse-Compton (EIC). The discussion in this section mainly follows \citet{2016MNRAS.460..396C}. All quantities in the jet comoving frame are denoted with a prime ($'$), and quantities without a prime are measured in the rest-frame of the BH (or host galaxy).

\subsubsection{Synchrotron Model}\label{sec:synchrotron-model}

At a distance $r=10^{14}r_{14}\,$cm from the BH, the magnetic field strength in the jet comoving frame is
\begin{equation}
\label{eq:B-field}
    B' = \left(2\xi_{\rm B}L_{\rm j} \over \Gamma^2 r^2 c \right)^{1/2} \simeq 8.2\times10^3\mr{\,G} {(L_{\rm j,48}\xi_{\rm B})^{1/2} \over r_{14} \Gamma_1}.
\end{equation}
The typical Lorentz factor of electrons radiating at $h\nu = \nu_{\rm keV}\,$keV frequency is given by
\begin{equation}
\label{eq:electron-gamma}
    \gamma'(\nu) = \left(4\pi \me c \nu \over 3\Gamma e B'
    \right)^{1/2} \simeq 7.6\times10^2 \left(\nu_{\rm keV}\over B_4'\Gamma_1 \right)^{1/2}\simeq 8.4\times10^2\, {(\nu_{\rm keV} r_{14})^{1/2} \over (L_{\rm j,48}\xi_{\rm B})^{1/4}}.
\end{equation}
The number of radiating electrons (or positrons) is directly given by the specific luminosity\footnote{It can be shown that synchrotron self-absorption by these electrons is negligible in the near-infrared, optical, and X-ray bands.},
\begin{equation}
\label{eq:electron_number}
\begin{split}
    \gamma'{\d N\over \gamma'} \simeq {L_\nu\over \Gamma} {\me c^2 \over e^3 B'}&\simeq 7.3\times10^{45}\, \nu_{\rm keV}^{-0.8} {L_{X,47}\over B_4' \Gamma_1}\\
    &\simeq 6.7\times10^{45}\, \gamma_3'^{-1.6} {L_{X,47} r_{14}^{1.8} \over (L_{\rm j, 48} \xi_{\rm B})^{0.9}},
\end{split}
\end{equation}
where we have made use of eqs. (\ref{eq:B-field}) and (\ref{eq:electron-gamma}) in the second row.

The kinetic energy carried by the X-ray emitting electrons is
\begin{equation}
    E_{\rm e} \simeq \Gamma \gamma' \me c^2 \gamma'{\d N\over \gamma'} \simeq
    6.1\times10^{43}\mr{\rm\,erg}\, 
    {L_{X,47} r_{14}^{3/2} \Gamma_1 \over \nu_{\rm keV}^{0.3} (L_{\rm j,48}\xi_{\rm B})^{3/4}},
\end{equation}
which is many orders of magnitude smaller than the X-ray energy radiated over a dynamical time. Thus, the radiating electrons are cooling off rapidly. This can also be seen by comparing their synchrotron cooling time with the dynamical time
\begin{equation}
\label{eq:t-cooling}
    {t_{\rm c}' \over t_{\rm dy}'} \simeq {6\pi \me c^2 \over \sigma_{\rm T} c B'^2 \gamma'} {\Gamma c\over r} \simeq 4.2\times10^{-5}\, {r_{14}^{1/2} \Gamma_1^3 \over \nu_{\rm keV}^{1/2} (L_{\rm j,48} \xi_{\rm B})^{3/4}}.
\end{equation}
However, there must be a mechanism preventing the accumulation of electrons with $\gamma'\ll 10^3$, because otherwise the near-infrared (K-band) flux will be over-produced. This argument applies to both SW1644 and SW2058. For the latter where dust extinction is weak, an extension of the X-ray flux according to the characteristic synchrotron power-law $L_\nu \propto \nu^{1/3}$ to the near-infrared band is only a factor of 2 below the observed flux (see the red dotted line in Figure \ref{fig:SED}). A possible mechanism preventing electrons' cooling is proposed by \citet{2015MNRAS.453.1820K} in the context of Poynting-dominated jet with $\sigma \gg 1$ and $\xi_{\rm B}\approx 1$. According to their model, the radiating electrons are re-accelerated and confined within magnetic reconnection regions (current sheets). At a given moment, the volume occupation fraction of active reconnection regions is roughly given by
\begin{equation}
    f_{\rm V}\sim {\sigma E_{\rm e}\over E_{\rm B}} \simeq 3.7\times10^{-6} \sigma\, {\eta_{X,-1} r_{14}^{1/2} \Gamma_1^3 \over \nu_{\rm keV}^{0.3} L_{\rm j,48}^{3/4}},
\end{equation}
where $E_{\rm B}\approx L_{\rm j} r/(2\Gamma^2 c)$ is the total magnetic energy within the causally connected region and we have used the X-ray radiation efficiency $\eta_{X} = L_X/L_{\rm j} = 0.1\eta_{X,-1}$. We note that a small instantaneous volume filling fraction $f_{\rm V}\ll 1$ does not necessarily mean a small overall radiative efficiency, because the reconnection regions may move around due to turbulence and hence sweep a much larger volume fraction over the dynamical time. Detailed modeling of the MHD kinematics is needed to further test this model.

On the high-frequency side, the data in the \textit{Swift} BAT $15(1+z)$--$50(1+z)\,$keV band in the first $\sim10\,$days is consistent with the extrapolation of the X-ray power-law \citep[][although we note the low photon statistics]{levan11}. Unfortunately, the SED in the MeV band is not strongly constrained, so we do not know where the power-law cuts off or breaks to a steeper segment.
Since electrons radiating above the X-ray band are in the fast cooling regime, a natural cutoff is the synchrotron burnoff limit at $\Gamma \me c^2/\alpha_{\rm FS}\sim 1\mr{\, GeV}\,(\Gamma/10)$ ($\alpha_{\rm FS}$ being the fine structure constant), corresponding to the maximum electron Lorentz factor at which the synchrotron cooling time is comparable to the Larmor timescale \citep{1992ApJ...396..161D, 2010ApJ...718L..63P}. Recent simulations of particle acceleration by relativistic magnetic reconnection, including radiation reaction force, show that this burnoff limit can be exceeded \citep{2014ApJ...782..104C, 2016ApJ...828...92Y}, because the electric field responsible for acceleration may be locally stronger than the magnetic field responsible for cooling. If the synchrotron spectrum indeed extends above $0.1(1+z)\,$MeV, then the \textit{Fermi} LAT upper limits are violated by at least two orders of magnitude \citep[as pointed out by][]{bloom11, burrows11}. In this picture, we require additional absorption of (sub-)GeV photons by $\gamma\gamma$ interaction with the X-rays.

Consider the propagation of a high-energy photon with frequency $\nu_h' = \nu_h/\Gamma$ through the causally connected length $r/\Gamma$ in the jet comoving frame. It primarily interacts with low-energy photons with frequency $\nu_\ell'$ given by
\begin{equation}
    h\nu_\ell' \simeq {4\me^2 c^4 \over h \nu_h'},
\end{equation}
and the peak cross-section is $\sigma_{\gamma\gamma}\simeq 0.2\sigma_{\rm T}$. The photon number density at $\nu_\ell'$ is given by
\begin{equation}
    n_{\rm ph}'(\nu_\ell')\simeq {\left.\nu L_{\rm \nu}\right|_{\nu_\ell} \over 4\pi r^2 c \Gamma^2 h\nu_\ell'}.
\end{equation}
Then the optical depth for $\gamma\gamma$ absorption is
\begin{equation}
\label{eq:taugg}
\begin{split}
    \tau_{\gamma\gamma}\simeq {n_{\rm ph}'(\nu_\ell') \sigma_{\gamma\gamma} r\over \Gamma} &\simeq 5.2\times10^2\, \nu_{\rm \ell,keV}^{-0.8} {L_{X,47} \over  r_{14} \Gamma_1^2}\\
&\simeq 2.5 \left(h\nu_h \over 0.1 (1+z)\mr{\,GeV}\right)^{0.8} {L_{X,47} \over r_{14} \Gamma_1^{3.6}}.
\end{split}
\end{equation}
We see that, in the synchrotron model, the non-detection of (sub-)GeV emission is expected if the jet Lorentz factor is modest $\Gamma\lesssim 13\, (L_{X,47}/r_{14})^{0.28}$.

The synchrotron seed photons will be inverse-Compton scattered by the same power-law electron distribution. Since the Compton-Y parameter is proportional to $\gamma'^3 \d N/\d \gamma'\propto \gamma'^{0.4}$, most of the synchrotron self-Compton (SSC) power is contributed by those electrons near the Klein-Nishina suppression threshold
\begin{equation}
    \gamma'_{\rm KN} \simeq {\Gamma \me c^2 \over h\nu_{\rm seed}},
\end{equation}
where $h\nu_{\rm seed}$ is the energy of the seed photons. Therefore, for a given seed-photon frequency $\nu_{\rm seed}$, the luminosity of the scattered photons is
\begin{equation}
\begin{split}
    \left(\nu L_\nu\right)_{\rm SSC} &\simeq {4\over 3} \left.\gamma'^3 {\d N\over \d \gamma'}\right|_{\gamma_{\rm KN}'} {\sigma_{\rm T}\over 4\pi r^2} \left.\nu L_\nu\right|_{\nu_{\rm seed}}\\
    &\simeq 2.1\times10^{45}\mr{\,erg\,s^{-1}}\, {L_{X,47}^2 \Gamma_1^{0.4}\over \nu_{\rm seed,keV}^{0.2} r_{14}^{0.2} (\xi_{\rm B} L_{\rm j,48})^{0.9}}.
\end{split}
\end{equation}
The typical energy of the scattered photons is $\Gamma \gamma_{\rm KN}' \me c^2\simeq 26\mr{\,GeV}\, \Gamma_1^2 \nu_{\rm seed,keV}^{-1}$, which is well within the energy range of \textit{Fermi} LAT. The strongest constraint was in the first few days when the X-ray flux reached $L_{\rm X}\sim10^{48}\rm\,erg\,s^{-1}$ during intensive flares. Taking a fixed X-ray radiative efficiency $\eta_{\rm X} = L_{\rm X}/L_{\rm j}$ and seed photon energy $h\nu_{\rm seed} = 0.5\,$keV, we can use the LAT limit $L_{\rm SSC}\lesssim 10^{46}\rm\,erg\,s^{-1}$ to obtain
\begin{equation}
{L_{\rm X}\over L_{\rm B}} = {\eta_{X}\over \xi_{\rm B}} \lesssim 2.8\times10^{-2}\, {r_{14}^{0.22}\over \Gamma_1^{0.44}}, 
\end{equation}
provided that GeV photons are not absorbed due to $\gamma$-$\gamma$ interaction. Therefore, in the synchrotron model, the non-detection of the SSC component in the (sub-)GeV band can also be explained if the X-ray radiative efficiency is low $\eta_{\rm X}\sim\,$a few percent and the jet is strongly magnetized $\xi_{\rm B}\sim 1$.

Finally, we conclude that the synchrotron emission remains a viable model for generating the X-rays from TDE jets. The \textit{Fermi} LAT (sub-)GeV limits may be explained by either $\gamma\gamma$ absorption in a low Lorentz-factor $\Gamma\lesssim 13\, (L_{X,47} /r_{14})^{0.28}$ jet or low X-ray radiative efficiency ($\eta_{X}\sim\,$a few percent). The lack of near-infrared (K-band) emission requires that the X-ray emitting electrons are re-accelerated on timescales shorter than their cooling time, which may be possible if electrons are confined within magnetic reconnection regions in a Poynting-dominated jet.

\subsubsection{Synchrotron Self-Compton (SSC) Model}
In this subsection, we discuss the model in which relativistic electrons inverse-Compton scatter their own synchrotron emission to the X-ray band. Over each scattering, the seed photons gain energy by a factor of $(4/3)\gamma_{\rm IC}'^2$, where $\gamma_{\rm IC}'$ is the Lorentz factor of electrons responsible for scattering. The probability of scattering is given by the Thomson optical depth of these electrons $\tau_{\rm IC}$. Thus, specific luminosity of seed photons at frequency $\nu_s$ is given by $L_{\nu_s}\simeq L_{\nu_X}/\tau_{\rm IC}\gg L_{\nu_X}$, provided that the emitting region is Thomson thin. Since the K-band ($h\nu\sim 1\,$eV) flux density is close to that in the X-ray band $F_{\nu_{\rm K}}\sim F_{\nu_{X}}\sim 0.1\rm\,mJy$, it is impossible to hide the synchrotron component between $\sim1\,$eV and the X-ray band without over-producing the K-band flux. Thus, the typical energy of seed photons must be $h\nu_s\ll 1\,$eV and hence the scattering electrons have typical Lorentz factor $\gamma_{\rm IC}'\gtrsim 30$. The K-band limit means the synchrotron luminosity is modest $L_{\rm syn}\lesssim 10^{44}\rm\,erg\,s^{-1}$. The luminosity of the scattered photons is given by $L_{X}\sim L_{\rm SSC} \sim L_{\rm syn}^2/L_{\rm B}$ (since the Compton-Y parameter is $Y\simeq L_{\rm syn}/L_{\rm B}$), which means that the jet magnetization is extremely weak $\xi_{\rm B} = L_{\rm B}/L_{\rm j}\sim L_{\rm syn}^2/(L_{\rm X}L_{\rm j})\lesssim 10^{-7} (L_{\rm X,47}L_{\rm j,48})^{-1}$.

A more serious problem than the weak magnetization is that the second-IC scattering generates high-energy photons in the \textit{Fermi} LAT band --- the maximum energy allowed by Klein-Nishina scattering is given by $\Gamma \gamma_{\rm IC}'\me c^2\gtrsim 0.15\mr{\,GeV}\, (\Gamma/10)$. It is possible to suppress (sub-)GeV flux by invoking a low-Lorentz factor jet $\Gamma<10$ such that the optical depth for $\gamma\gamma$ pair production $\tau_{\gamma\gamma}\gg 1$ (see eq. \ref{eq:taugg}). However, these pairs carries a large amount of energy $L_{e^\pm} \sim YL_{\rm X}$, which clearly exceeds the energy budget of the system ($10^{53}$---$10^{54}\,$erg).

We conclude that SSC emission is not a viable mechanism for generating the X-rays from TDE jets. A more comprehensive study of the full parameter space is presented by \citet{2016MNRAS.460..396C}, where the authors show that the SSC scenario cannot avoid an excess at a different wavelength or requiring too much energy.

\subsubsection{External inverse-Compton (EIC) Model}
In this subsection, we discuss the possibility that the X-rays are generated by jet electrons inverse-Compton scattering soft photons from external sources in the ambiemt medium. TDEs typically have very large optical-UV photospheric radii of order $10^{14}$---$10^{15}\,$cm inferred from the blackbody fit to their SEDs. The optical-UV emission from SW1644 was extinguished by a large dust column, but the thermal UV component was clearly seen in SW2058 \citep{cenko12, pasham15} where the inferred photospheric radius is $r_{\rm ph}\sim\,$a few$ \times10^{14}\,$cm. We note that $r_{\rm ph}$ inferred from blackbody fits are not the radius where the scattering optical depth equals to unity. If the optical-UV radiation is generated by reprocessing of higher-energy photons by an outflow with density profile $\rho\propto r^{-2}$ \citep{strubbe09, metzger16, roth16}, then photons diffuse away at the trapping radius $r_{\rm tr}$ where the scattering optical depth becomes $c/v_{\rm w}$ ($v_{\rm w}$ being the outflow speed). If the radiation energy density at the trapping radius is denoted as $U_{\rm ph}$, then the photospheric radius inferred from $L_{\rm uv} \simeq 4\pi r_{\rm tr}^2 U_{\rm uv}v_{\rm w} \simeq 4\pi r_{\rm ph}^2 \sigma_{\rm SB} T^4$ is $r_{\rm ph}\simeq r_{\rm tr}\sqrt{4v_{\rm w}/c}$.

As the jet goes through the optical-UV radiation bath, electrons will scatter ambient photons to higher energies
\begin{equation}
    h\nu_{\rm IC} \sim \Gamma^2\gamma'^2 h\nu_{\rm uv}.
\end{equation}
We see that, for a jet Lorentz factor $\Gamma\sim 10$, cold or mildly relativistic electrons/positrons $\gamma'\sim 1$ are capable of producing the X-rays ($h\nu_{\rm IC}\sim 1\,$keV) by scattering UV photons ($h\nu_{\rm uv}\sim 10\,$eV). In the following, we assume cold leptons in the jet comoving frame, based on the argument of fast-cooling (eq. \ref{eq:t-cooling}).

If the number of leptons per proton mass is $\xi_{\rm e}\in [1, \mp/\me]$ and the magnetization is $\sigma$, then Thomson optical depth of the jet at radius $r$ is given by
\begin{equation}
    \tau_{\rm j} \simeq {\sigma_{\rm T}L_{\rm j}\xi_{\rm e}/(1+\sigma) \over 4\pi r \Gamma^3 \mp c^3} \simeq 1.17\times10^{-2} {L_{\rm j,48} \xi_{\rm e} \over r_{14} \Gamma_1^3 (1+\sigma)}.
\end{equation}
The radiation energy density near the photospheric radius $r_{\rm ph}$ is given by $U_{\rm uv}\simeq 4L_{\rm uv}/(4\pi r_{\rm ph}^2 c)$. If the jet is optically thin, the inverse-Compton luminosity of each electrons is given by $\sim\sigma_{\rm T} c U_{\rm uv} \Gamma^4$, where an additional factor of $\Gamma^2$ is because the electron is moving towards the observer at a relativistic speed. The total number of electrons in the causally connected region is given by $N_{\rm e} = 4\pi r^2 \tau_{\rm j}/\sigma_{\rm T}$. Thus, the total luminosity due to EIC scattering near radius $r_{\rm ph}$ is given by
\begin{equation}
    L_{\rm EIC} \simeq 4\tau_{\rm j} \Gamma^4 L_{\rm uv}\simeq 
    (4.7\times10^{47}\mr{\,erg\,s^{-1}})\, L_{\rm uv,45} {L_{\rm j,48}\Gamma_1 \xi_{\rm e} \over r_{\rm ph,14} (1+\sigma)}.
\end{equation}
We see that the EIC scattering by a relativistic cold baryonic jet ($\xi_{\rm e}\sim 1$ and $\sigma\ll 1$) naturally reproduces the photon energy and luminosity of the X-ray emission \citep[as shown by][]{burrows11, lu17}. We note that the strength of the inverse-Compton drag is given by $\eta_{\rm drag} = L_{\rm EIC}/L_{\rm j}$, which may reach order unity for a baryonic jet going through the ambient radiation bath of $L_{\rm uv}\sim 10^{45}\rm\,erg\,s^{-1}$ and $r_{\rm ph}\sim10^{14}\,$cm.

The difficulty of the EIC model is to generate the power-law X-ray spectrum. This is because strong inverse-Compton cooling (similar as in eq. \ref{eq:t-cooling}) quickly removes any power-law tail at $\gamma'\gg1$, unless additional re-acceleration is invoked. It has been proposed that repetitive scattering in a flow with strong shear motion can produce a power-law spectrum in a Fermi-like process \citep{lu17}. This scenario relies on the jet structure in the transverse direction such that a fraction of the photons scattered near the jet axis are reflected back by the funnel wall and may get scattered again. If the jet has opening angle $\theta_{\rm j}\sim \Gamma_{\rm j}^{-1}$, the optical depth of the jet in the transverse direction is given by $\tau_{\rm trvs}\simeq \tau_{\rm j}\Gamma$. If the reflection fraction is $\eta_{\rm ref}$, the Compton-y parameter for repetitive scattering in a baryonic jet ($\xi_{\rm e}\sim 1$ and $\sigma\ll 1$) is given by
\begin{equation}
    y\sim \Gamma^2 \tau_{\rm trvs}\eta_{\rm ref} \sim 1.2\, {\eta_{\rm ref}\over 0.1}{L_{\rm j,48} \over r_{\rm ph,14}}.
\end{equation}
Note that the Compton-y parameter is independent of the jet Lorentz factor, which means all relativistically moving regions (at different angles wrt. the jet axis) in a structured jet make contributions proportional to their local isotropic power $L_{\rm j}$. Thus, repetitive scattering by a cold jet may generate a power-law spectrum through bulk Comptonization, provided that the jet is surrounded by a reflector which constantly reflects $\sim10\%$ of the scattered photons. General relativistic radiative MHD simulations of jetted TDEs by \citet[][their Figure 12]{curd19} show that the bulk Comptonization model is in qualitative agreement with the X-ray luminosity and spectrum for SW1644, when the observer is looking down along the jet axis.

A prediction of this model is that the power-law spectrum should cut off at $\sim5\mr{\,MeV}\,(\Gamma/10)$, because the scattering cross-section is suppressed when photon energies in the jet comoving frame approach the Klein-Nishina regime $h\nu'\sim \me c^2$. This is distinguishable from the synchrotron model in subsection \ref{sec:synchrotron-model} where the power-law should extends up to 10's or 100's MeV (when $\gamma\gamma$ absorption becomes important, see eq. \ref{eq:taugg}).


\section{Multi-Messenger Astronomy}
\label{sec:multi}

In this section, we briefly discuss the multi-messenger aspects of TDEs, including gravitational waves (GWs), PeV neutrinos and ultra-high energy cosmic rays (UHECRs).

(1) A star in a parabolic orbit will plunge into the event horizon as a whole if the pericenter radius $r_{\rm p}$ is less than twice the Schwarzschild radius of the BH, for a non-spinning BH \citep[the maximum mass is larger for rapidly spinning BHs,][]{east14}. For the marginal disruption case ($r_{\rm p}\simeq r_{\rm T} \simeq 2R_{\rm S}$) at luminosity distance $D_{\rm L}$, the peak GW strain is $h_{\rm max}\sim {GM_\star/ (2D_{\rm L}c^2)}\sim 2\times10^{-20}\, (M_\star/\msun)(D_{\rm L}/1\rm\,Mpc)^{-1}$. However, most TDEs have tidal disruption radius larger than $2R_{\rm S}$, so the corresponding peak strain is reduced by a factor of $2R_{\rm S}/r_{\rm T}\sim 0.1\,M_6^{2/3}M_{\rm\star,\odot}^{1/3}R_{\rm \star,\odot}^{-1}$. The peak GW frequency for a TDE is $f_{\rm peak}\sim (2\pi)^{-1} \sqrt{2GM_\star/R_\star^3}$, which is $\sim 0.1\mr{\,Hz}\,(M_\star/0.6\msun)$ for white dwarfs and $\sim 4\times10^{-4}\mr{\,Hz}\, (M_\star/\msun)^{-1}$ for main-sequence stars. These TDEs may be detected by LISA as short GW bursts lasting for a timescale of $(2\pi f_{\rm peak})^{-1}$ \citep{2004ApJ...615..855K, 2009ApJ...695..404R}. Taking a LISA sensitivity of $\sim10^{-20}$ around $10^{-3}\,$Hz and $0.1\,$Hz \citep{2018arXiv180301944R}, we see that only TDEs within the Local Group may be detected in GW.

(2) Jetted TDEs have been suggested as possible sources of high-energy diffuse (sub-)PeV neutrinoes \citep{wang16}. This model can
avoid overproducing the diffuse $\gamma$-ray background measured by \textit{Fermi} \citep{2016PhRvL.116g1101M}, since GeV-TeV photons from pion decay are absorbed when interacting with the X-ray radiation field (eq. \ref{eq:taugg}). The (sub-)PeV neutrino flux of $10^{-7}\rm \,GeV\,cm^{-2} \,s^{-1}\,sr^{-1}$ requires a cosmic ray (CR) energy injection rate of $\sim10^{45}\rm\,erg\,Mpc^{-3}\,yr^{-1}$ \citep{dai17, lunardini17}. Taking a bolometric correction of a factor of $\sim3$, the isotropic equivalent electromagnetic (EM) energy injection for each of the three \textit{Swift} events is of order $10^{54}\,$erg. Inferred from \textit{Swift} observations \citep{{burrows11, brown15}}, the rate of jetted TDEs whose $\gamma$-ray emission are beamed towards the Earth is $\sim 3\times10^{-11} \rm\,Mpc^{-3} \,yr^{-1}$, which gives an EM energy injection rate $3\times10^{43}\rm \,erg\,Mpc^{-3}\,yr^{-1}$ that is a factor of 30 less than the required CR injection rate. We see that jetted TDEs may be sub-dominant neutrino sources. On the other hand, this discrepancy may be reconciled by a very low bolometric radiative efficiency $\lesssim 3\%$ (and nearly all the jet energy goes into CRs). Another possibility is that many weak TDE jets are hydrodynamically choked or Compton dragged to low Lorentz factors such that their $\gamma$-ray emission is strongly suppressed (and hence undetectable by \textit{Swift}). The next generation of all-sky soft X-ray surveys (e.g. eROSITA and Einstein Probe) should provide a more accurate jetted TDE rate.


(3) Jetted TDEs have also been suggested as possible sources of UHECRs \citep{farrar09, 2014arXiv1411.0704F}. According to the Hillas criteion \citep{1984ARA&A..22..425H}, the maximum CR energy in the jet comoving frame $\mathcal{E}'$ is limited by the Larmor radius being less than the size of causally connected region $r/\Gamma$, i.e. $\mathcal{E}'\sim Ze B'r/\Gamma$, where $B'$ is the B-field strength and $Ze$ is the charge of the particle. For a Poynting jet with isotropic equivalent luminosity $L_{\rm j}$, we have $B'r\sim \sqrt{L_{\rm j}/\Gamma^2 c}$ and hence the maximum CR energy is $\mathcal{E}\simeq \Gamma \mathcal{E}'\sim 10^{20}\mr{\,eV}\,Z L_{\rm j,48}^{1/2}/\Gamma_1$. To explain the bulk UHECR flux, a CR energy injection rate similar to that required by the high-energy neutrinoes is needed. Again, the debate comes down to the uncertain rate of jetted TDEs.

\section{Conclusion}
\label{sec:concl}

Jetted TDEs are the youngest members of the family of relativistic jets. These events provide an ideal testbed to study many aspects of jet physics:
\begin{itemize}
    \item[1.] The sudden switch-on and then switch-off over a period of two years in SW1644 and SW2058 indicated that the jet launching may be associated with super-Eddington accretion.
    \item[2.] The occurrence rate of the \textit{Swift} events constrains the fraction of TDEs with relativistic jets to be $f_{\rm j}\gtrsim 3\times 10^{-3}$. On the other hand, the absence of strong radio emission from very nearby TDEs (e.g. ASASSN-14li) shows that jet launching is not universal in all TDEs, which implies that super-Eddington accretion is not a sufficient condition for launching relativistic jets.
    \item[3.] The radio afterglow of SW1644 indicates that the total jet kinetic energy is of order $10^{53}$~erg, which means that the jet launching process must be highly efficient at least for this event (given the total energy budget of $10^{54}$~erg).
    \item[4.] The multi-wavelength data, in particular the lack of near-infrared and GeV emission, from SW1644 and SW2058 put interesting constraints on the $\gamma/$X-ray emission mechanism and jet composition. For instance, the synchrotron self-Compton model has been ruled out for the X-ray emission from these events. The two surviving models are synchrotron emission (which requires a Poynting-dominated jet) and external Compton scattering (off thermal UV-optical photons).
\end{itemize}

Despite the delightful progress above, much more work is still needed in this young research field.
\begin{itemize}
    \item[1.] The dynamics of the jet (determining the observed emission), depending on a the large variety of TDE conditions (i.e., different stellar and black hole masses, impact parameters, the magnetic field, etc) and on the SMBH environment, has been only partically studied.
    \item[2.] In particular, with respect to the dynamics of the jet through the very close-by region, polluted by the TDE debris, the unanswered questions are: How does the TDE debris (and Compton drag by UV-optical photons) affect the dynamics of baryonic/leptonic/Poynting dominated jets? What are the signatures of the extended cocoon generated by the jet-environment interaction? How does Lense-Thirring precession (due to misalignment between disk angular momentum and BH spin) affect the jet propagation?
    \item[3.] The biggest limit to our understanding of jetted TDEs is the low number of detections (and hence the jet fraction remains uncertain by several orders of magnitude). The unanswered questions are: What is the most promising way of finding jetted TDEs? Is it possible to observe the thermal emission and afterglow from the cocoon (specially in relativistic jets seen off-axis)? Are there more $\gamma/$X-ray bright events buried in \textit{Swift} archive data? What is the typical angular structure of TDE jets?
    \item[4.] It is unclear why some TDEs produce relativistic jets (as indicated by bright $\gamma/$X-ray and radio emission) while others only have non-relativistic outflows. Also, what processes control the on-and-off switch of jets in the \textit{Swift} events?
    \item[5.] It is still unclear why the jets in the \textit{Swift} events pointed towards the observer in a quasi-steady manner, given that Lense-Thirring precession should quickly cause the narrowly beamed jet emission to move out of our line of sight. If the jet axis was brought to alignment with the BH spin, how was the alignment process achieved? What are the signatures of Lense-Thirring precession of TDE jets?
    \item[6.] TDE jets may be interesting multi-messenger sources (of PeV neutrinos and UHECRs), given that current observations allow a jet fraction as high as $f_{\rm j}\sim 0.1$.
\end{itemize}

Looking into the near future, the sample of jetted TDEs should grow by one or two orders of magnitude. Wide field-of-view radio transient surveys by ASKAP/VAST and VLASS will likely find $\sim$10 off-axis jetted TDEs per year \citep{metzger15}. The ngVLA survey at 3--100~GHz will likely find $\sim$100 on-axis jetted TDEs per year \citep{vanvelzen18}. Moreover, follow-up observations of optically selected TDEs by ngVLA will be able to detect faint radio emission at the level of ASASSN-14li up to $z=0.2$ \citep{vanvelzen18}, and hence the population of off-axis jetted TDEs will be strongly constrained. On the other hand, all-sky X-ray surveys eROSITA \citep{2012arXiv1209.3114M} and Einstein Probe \citep{2015arXiv150607735Y} may be able to find a few on-axis jetted TDEs per year without \textit{Swift} $\gamma$-ray triggers \citep{2014MNRAS.437..327K}. These next generation of facilities will shed light on many of the open questions.

\vskip 1cm
\section*{Acknowledgements}

FDC  acknowledges  support  from  the  UNAM-PAPIIT grant IN117917. WL was supported by the David and Ellen Lee Fellowship at Caltech.


\begin{thebibliography}{00}

\bibitem[Abramowicz \& Liu(2012)]{abramowicz12} Abramowicz, M.~A., \& Liu, F.~K.\ 2012, \aap, 548, A3 

\bibitem[Aleksi{\'c} et al.(2013)]{aleksic13} Aleksi{\'c}, J., Antonelli, L.~A., Antoranz, P., et al.\ 2013, \aap, 552, A112 

\bibitem[Alexander et al.(2016)]{alexander16} Alexander, K.~D., Berger, E., Guillochon, J., Zauderer, B.~A., \& Williams, P.~K.~G.\ 2016, \apjl, 819, L25 

\bibitem[Alexander et al.(2017)]{alexander17} Alexander, K.~D., Wieringa, M.~H., Berger, E., Saxton, R.~D., \& Komossa, S.\ 2017, \apj, 837, 153 

\bibitem[Aliu et al.(2011)]{aliu11} Aliu, E., Arlen, T., Aune, T., et al.\ 2011, \apjl, 738, L30 

\bibitem[Arcavi et al.(2014)]{arcavi14} Arcavi, I., Gal-Yam, A., Sullivan, M., et al.\ 2014, \apj, 793, 38 

\bibitem[Auchettl et al.(2017)]{auchettl17} Auchettl, K., Guillochon, J., \& Ramirez-Ruiz, E.\ 2017, \apj, 838, 149 

\bibitem[Auchettl et al.(2018)]{auchettl18} Auchettl, K., Ramirez-Ruiz, E., \& Guillochon, J.\ 2018, \apj, 852, 37 

\bibitem[Bade et al.(1996)]{bade96} Bade, N., Komossa, S., \& Dahlem, M.\ 1996, \aap, 309, L35 

\bibitem[Barniol Duran \& Piran(2013)]{barniolduran13} Barniol Duran, R., \& Piran, T.\ 2013, \apj, 770, 146 

\bibitem[Barres de Almeida \& De Angelis(2011)]{barres11} Barres de Almeida, U., \& De Angelis, A.\ 2011, arXiv:1104.2528 

\bibitem[Begelman \& Cioffi(1989)]{begelman89} Begelman, M.~C., \& Cioffi, D.~F.\ 1989, \apjl, 345, L21 

\bibitem[Berger et al.(2012)]{berger12} Berger, E., Zauderer, A., Pooley, G.~G., et al.\ 2012, \apj, 748, 36.

\bibitem[Blagorodnova et al.(2017)]{blagorodnova17} Blagorodnova, N., Gezari, S., Hung, T., et al.\ 2017, \apj, 844, 46 

\bibitem[Blandford \& McKee(1976)]{blandford76} Blandford, R.~D., \& McKee, C.~F.\ 1976, Physics of Fluids, 19, 1130 

\bibitem[Blandford \& Znajek(1977)]{blandford77} Blandford, R.~D., \& Znajek, R.~L.\ 1977, \mnras, 179, 433 

\bibitem[Bloom et al.(2011)]{bloom11} Bloom, J.~S., Giannios, D., Metzger, B.~D., et al.\ 2011, Science, 333, 203.

\bibitem[Bonnerot et al.(2016)]{2016MNRAS.455.2253B} Bonnerot, C., Rossi, E.~M., Lodato, G., \& Price, D.~J.\ 2016, \mnras, 455, 2253 

\bibitem[Bonnerot et al.(2017)]{bonnerot17} Bonnerot, C., Price, D.~J., Lodato, G., \& Rossi, E.~M.\ 2017, \mnras, 469, 4879 

\bibitem[Bower et al.(2013)]{bower13} Bower, G.~C., Metzger, B.~D., Cenko, S.~B., Silverman, J.~M., \& Bloom, J.~S.\ 2013, \apj, 763, 84 

\bibitem[Bright et al.(2018)]{bright18} Bright, J.~S., Fender, R.~P., Motta, S.~E., et al.\ 2018, \mnras, 475, 4011 

\bibitem[Brown et al.(2015)]{brown15} Brown, G.~C., Levan, A.~J., Stanway, E.~R., et al.\ 2015, \mnras, 452, 4297 

\bibitem[Brown et al.(2017)]{brown17} Brown, G.~C., Levan, A.~J., Stanway, E.~R., et al.\ 2017, \mnras, 472, 4469 

\bibitem[Brown et al.(2017)]{brown18} Brown, J.~S., Holoien, T.~W.-S., Auchettl, K., et al.\ 2017, \mnras, 466, 4904 


\bibitem[Burrows et al.(2011)]{burrows11} Burrows, D.~N., Kennea, J.~A., Ghisellini, G., et al.\ 2011, \nat, 476, 421 

\bibitem[Cao \& Wang(2012)]{cao12} Cao, D., \& Wang, X.-Y.\ 2012, \apj, 761, 111 

\bibitem[Cendes et al.(2014)]{cendes14} Cendes, Y., Wijers, R.~A.~M.~J., Swinbank, J.~D., et al.\ 2014, arXiv:1412.3986 

\bibitem[Cenko et al.(2012)]{cenko12} Cenko, S.~B., Krimm, H.~A., Horesh, A., et al.\ 2012, \apj, 753, 77 

\bibitem[Cerutti et al.(2014)]{2014ApJ...782..104C} Cerutti, B., Werner, G.~R., Uzdensky, D.~A., \& Begelman, M.~C.\ 2014, \apj, 782, 104 

\bibitem[Cheng et al.(2016)]{cheng16} Cheng, K.~S., Chernyshov, D.~O., Dogiel, V.~A., Kong, A.~K.~H., \& Ko, C.~M.\ 2016, \apjl, 816, L10 

\bibitem[Chornock et al.(2014)]{chornock14} Chornock, R., Berger, E., Gezari, S., et al.\ 2014, \apj, 780, 44 


\bibitem[Crumley et al.(2016)]{2016MNRAS.460..396C} Crumley P., Lu W., Santana R., Hern{\'a}ndez R.~A., Kumar P., Markoff S., 2016, MNRAS, 460, 396

\bibitem[Curd \& Narayan(2019)]{curd19} Curd, B., \& Narayan, R.\ 2019, \mnras, 483, 565 

\bibitem[Dai \& Fang(2017)]{dai17} Dai, L., \& Fang, K.\ 2017, \mnras, 469, 1354 

\bibitem[De Colle et al.(2012)]{decolle12} De Colle, F., Guillochon, J., Naiman, J., \& Ramirez-Ruiz, E.\ 2012a, \apj, 760, 103 

\bibitem[de Jager \& Harding(1992)]{1992ApJ...396..161D} de Jager, O.~C., \& Harding, A.~K.\ 1992, \apj, 396, 161 

\bibitem[Donnarumma \& Rossi(2015)]{donnarumma15} Donnarumma, I., \& Rossi, E.~M.\ 2015, \apj, 803, 36 



\bibitem[East(2014)]{east14} East, W.~E.\ 2014, \apj, 795, 135 

\bibitem[Eftekhari et al.(2018)]{eftekhari18} Eftekhari, T., Berger, E., Zauderer, B.~A., Margutti, R., \& Alexander, K.~D.\ 2018, \apj, 854, 86 

\bibitem[Evans \& Kochanek(1989)]{evans89} Evans, C.~R., \& Kochanek, C.~S.\ 1989, \apjl, 346, L13 

\bibitem[Farrar \& Gruzinov(2009)]{farrar09} Farrar, G.~R., \& Gruzinov, A.\ 2009, \apj, 693, 329 

\bibitem[Farrar \& Piran(2014)]{2014arXiv1411.0704F} Farrar, G.~R., \& Piran, T.\ 2014, arXiv:1411.0704 


\bibitem[Fossati et al.(1998)]{fossati98} Fossati, G., Maraschi, L., Celotti, A., Comastri, A., \& Ghisellini, G.\ 1998, \mnras, 299, 433 


\bibitem[Generozov et al.(2017)]{generozov17} Generozov, A., Mimica, P., Metzger, B.~D., et al.\ 2017, \mnras, 464, 2481 

\bibitem[Gezari et al.(2008)]{2008ApJ...676..944G} Gezari, S., Basa,
  S., Martin, D.~C., et al.\ 2008, \apj, 676, 944-969 

\bibitem[Gezari et al.(2009)]{2009ApJ...698.1367G} Gezari, S.,  Heckman, T., Cenko, S.~B., et al.\ 2009, \apj, 698, 1367 

\bibitem[Gezari et al.(2012)]{2012Natur.485..217G} Gezari, S.,  Chornock, R., Rest, A., et al.\ 2012, \nat, 485, 217  

\bibitem[Ghisellini et al.(2017)]{ghisellini17} Ghisellini, G., Righi, C., Costamante, L., \& Tavecchio, F.\ 2017, \mnras, 469, 255 


\bibitem[Giannios \& Metzger(2011)]{giannios11} Giannios D., Metzger B.~D., 2011, MNRAS, 416, 2102

\bibitem[Granot et al.(2002)]{granot02} Granot, J., Panaitescu, A., Kumar, P., \& Woosley, S.~E.\ 2002, \apjl, 570, L61 

\bibitem[Granot \& Sari(2002)]{2002ApJ...568..820G} Granot, J., \& Sari, R.\ 2002, \apj, 568, 820 

\bibitem[Granot \& Piran(2012)]{2012MNRAS.421..570G} Granot, J., \& Piran, T.\ 2012, \mnras, 421, 570 

\bibitem[Grupe et al.(1999)]{grupe99} Grupe, D., Thomas, H.-C., \& Leighly, K.~M.\ 1999, \aap, 350, L31 


\bibitem[Guillochon \& Ramirez-Ruiz(2013)]{guillochon13} Guillochon, J., \& Ramirez-Ruiz, E.\ 2013, \apj, 767, 25 

\bibitem[Guillochon \& McCourt(2017)]{guillochon17} Guillochon, J., \& McCourt, M.\ 2017, \apjl, 834, L19 


\bibitem[\protect\citeauthoryear{Hillas}{1984}]{1984ARA&A..22..425H} Hillas A.~M., 1984, ARA\&A, 22, 425

\bibitem[Hills(1975)]{hills75} Hills, J.~G., 1975, \nat, 254, 295


\bibitem[Holoien et al.(2016)]{holoien16} Holoien, T.~W.-S., Kochanek, C.~S., Prieto, J.~L., et al.\ 2016, \mnras, 455, 2918 

\bibitem[Irwin et al.(2015)]{irwin15} Irwin, J.~A., Henriksen, R.~N., Krause, M., et al.\ 2015, \apj, 809, 172 

\bibitem[Jiang et al.(2016a)]{2016ApJ...830..125J} Jiang, Y.-F., Guillochon, J., \& Loeb, A.\ 2016a, \apj, 830, 125 

\bibitem[Jiang et al.(2016b)]{2016ApJ...828L..14J} Jiang, N., Dou, L., Wang, T., et al.\ 2016b, \apjl, 828, L14 

\bibitem[Kara et al.(2018)]{kara18} Kara, E., Dai, L., Reynolds, C.~S., \& Kallman, T.\ 2018, \mnras, 474, 3593 

\bibitem[Kawamuro et al.(2016)]{kawamuro16} Kawamuro, T., Ueda, Y., Shidatsu, M., et al.\ 2016, \pasj, 68, 58 


\bibitem[Kelley et al.(2014)]{kelley14} Kelley, L.~Z., Tchekhovskoy, A., \& Narayan, R.\ 2014, \mnras, 445, 3919 


\bibitem[Khabibullin et al.(2014)]{2014MNRAS.437..327K} Khabibullin, I., Sazonov, S., \& Sunyaev, R.\ 2014, \mnras, 437, 327 

\bibitem[Kobayashi et al.(2004)]{2004ApJ...615..855K} Kobayashi, S., Laguna, P., Phinney, E.~S., \& M{\'e}sz{\'a}ros, P.\ 2004, \apj, 615, 855 

\bibitem[Komossa \& Bade(1999)]{komossa99} Komossa, S., \& Bade, N.\ 1999, \aap, 343, 775 

\bibitem[Komossa(2002)]{komossa02} Komossa, S.\ 2002, Lighthouses of the Universe: The Most Luminous Celestial Objects and Their Use for Cosmology, 436 

\bibitem[Komossa(2015)]{komossa15} Komossa, S.\ 2015, Journal of High Energy Astrophysics, 7, 148 

\bibitem[Krolik \& Piran(2011)]{krolik11} Krolik, J.~H., \& Piran, T.\ 2011, \apj, 743, 134 


\bibitem[Krolik et al.(2016)]{krolik16} Krolik, J., Piran, T., Svirski, G., \& Cheng, R.~M.\ 2016, \apj, 827, 127 

\bibitem[Kumar et al.(2013)]{kumar13} Kumar, P., Barniol Duran, R., Bo{\v s}njak, {\v Z}., \& Piran, T.\ 2013, \mnras, 434, 3078 

\bibitem[Kumar \& Zhang(2015)]{kumar15} Kumar, P., \& Zhang, B.\ 2015, \physrep, 561, 1.

\bibitem[Kumar \& Crumley(2015)]{2015MNRAS.453.1820K} Kumar P., Crumley P., 2015, MNRAS, 453, 1820

\bibitem[Lacy et al.(1982)]{1982ApJ...262..120L} Lacy, J.~H., Townes, C.~H., \& Hollenbach, D.~J.\ 1982, \apj, 262, 120 



\bibitem[Lei et al.(2016)]{lei16} Lei, W.-H., Yuan, Q., Zhang, B., \& Wang, D.\ 2016, \apj, 816, 20 



\bibitem[Levan et al.(2011)]{levan11} Levan, A.~J., Tanvir, N.~R., Cenko, S.~B., et al.\ 2011, Science, 333, 199 

\bibitem[Levan et al.(2014)]{levan14} Levan, A.~J., Tanvir, N.~R., Starling, R.~L.~C., et al.\ 2014, \apj, 781, 13 

\bibitem[Levan et al.(2016)]{levan16} Levan, A.~J., Tanvir, N.~R., Brown, G.~C., et al.\ 2016, \apj, 819, 51 


\bibitem[Liska et al.(2018)]{liska18} Liska, M.~T.~P., Tchekhovskoy, A., \& Quataert, E.\ 2018, arXiv e-prints , arXiv:1809.04608.

\bibitem[Lodato \& Rossi(2011)]{2011MNRAS.410..359L} Lodato, G., \& Rossi, E.~M.\ 2011, \mnras, 410, 359 


\bibitem[Loeb \& Ulmer(1997)]{loeb97} Loeb, A., \& Ulmer, A.\ 1997, \apj, 489, 573 

\bibitem[Lu et al.(2016)]{lu16} Lu, W., Kumar, P., \& Evans, N.~J.\ 2016, \mnras, 458, 575 

\bibitem[Lu et al.(2017)]{lu17} Lu, W., Krolik, J., Crumley, P., \& Kumar, P.\ 2017, \mnras, 471, 1141 

\bibitem[Lu \& Kumar(2018)]{lu18} Lu, W., \& Kumar, P.\ 2018, \apj, 865, 128 

\bibitem[Lunardini \& Winter(2017)]{lunardini17} Lunardini, C., \& Winter, W.\ 2017, \prd, 95, 123001 

\bibitem[Mangano et al.(2016)]{mangano16} Mangano, V., Burrows, D.~N., Sbarufatti, B., \& Cannizzo, J.~K.\ 2016, \apj, 817, 103 

\bibitem[Mattila et al.(2018)]{mattila18} Mattila, S., P{\'e}rez-Torres, M., Efstathiou, A., et al.\ 2018, Science, 361, 482 

\bibitem[Merloni et al.(2012)]{2012arXiv1209.3114M} Merloni, A., Predehl, P., Becker, W., et al.\ 2012, arXiv:1209.3114 

\bibitem[Metzger et al.(2012)]{metzger12} Metzger, B.~D., Giannios, D., \& Mimica, P.\ 2012, \mnras, 420, 3528 

\bibitem[Metzger et al.(2015)]{metzger15} Metzger, B.~D., Williams, P.~K.~G., \& Berger, E.\ 2015, \apj, 806, 224 

\bibitem[Metzger \& Stone(2016)]{metzger16} Metzger, B.~D., \& Stone, N.~C.\ 2016, \mnras, 461, 948 


\bibitem[Mimica et al.(2015)]{mimica15} Mimica, P., Giannios, D., Metzger, B.~D., \& Aloy, M.~A.\ 2015, \mnras, 450, 2824 

\bibitem[\protect\citeauthoryear{Murase, Guetta \& Ahlers}{2016}]{2016PhRvL.116g1101M} Murase K., Guetta D., Ahlers M., 2016, PhRvL, 116, 71101

\bibitem[Niko{\l}ajuk \& Walter(2013)]{nikolajuk13} Niko{\l}ajuk, M., \& Walter, R.\ 2013, \aap, 552, A75 

\bibitem[Omodei et al.(2011)]{omodei11} Omodei, N., Troja, E., Corbet, R., Perkins, J.~S., \& McEnery, J.~E.\ 2011, GRB Coordinates Network, Circular Service, No.~11862, \#1 (2011), 11862, 1 



\bibitem[Pasham et al.(2015)]{pasham15} Pasham, D.~R., Cenko, S.~B., Levan, A.~J., et al.\ 2015, \apj, 805, 68 

\bibitem[Pasham \& van Velzen(2018)]{pasham18} Pasham, D.~R., \& van Velzen, S.\ 2018, \apj, 856, 1

\bibitem[Peng et al.(2016)]{peng16} Peng, F.-K., Tang, Q.-W., \& Wang, X.-Y.\ 2016, \apj, 825, 47.

\bibitem[Perlman et al.(2017)]{perlman17} Perlman, E.~S., Meyer, E.~T., Wang, Q.~D., et al.\ 2017, \apj, 842, 126 

\bibitem[Phinney(1989)]{1989IAUS..136..543P} Phinney, E.~S.\ 1989, The Center of the Galaxy, 136, 543  

\bibitem[Piran \& Nakar(2010)]{2010ApJ...718L..63P} Piran, T., \& Nakar, E.\ 2010, \apjl, 718, L63 

\bibitem[Quataert(2004)]{quataert04} Quataert, E.\ 2004, \apj, 613, 322 

\bibitem[Quataert \& Kasen(2012)]{quataert12} Quataert, E., \& Kasen, D.\ 2012, \mnras, 419, L1 

\bibitem[Rees(1988)]{rees88} Rees, M.~J., 1988, \nat, 333, 523 

\bibitem[Reis et al.(2012)]{reis12} Reis, R.~C., Miller, J.~M., Reynolds, M.~T., et al.\ 2012, Science, 337, 949 

\bibitem[Robson et al.(2018)]{2018arXiv180301944R} Robson T., Cornish N., Liu C., 2018, arXiv e-prints, arXiv:1803.01944

\bibitem[Romero-Ca{\~n}izales et al.(2016)]{romero16} Romero-Ca{\~n}izales, C., Prieto, J.~L., Chen, X., et al.\ 2016, \apjl, 832, L10 

\bibitem[Rosswog et al.(2009)]{2009ApJ...695..404R} Rosswog, S., Ramirez-Ruiz, E., \& Hix, W.~R.\ 2009, \apj, 695, 404 

\bibitem[Roth et al.(2016)]{roth16} Roth, N., Kasen, D., Guillochon, J., \& Ramirez-Ruiz, E.\ 2016, \apj, 827, 3 

\bibitem[S{\c a}dowski et al.(2016)]{2016MNRAS.458.4250S} S{\c a}dowski, A., Tejeda, E., Gafton, E., et al.\ 2016, \mnras, 458, 4250.

\bibitem[Sari \& Piran(1995)]{sari95} Sari, R., \& Piran, T.\ 1995, \apjl, 455, L143 

\bibitem[Saxton et al.(2012)]{saxton12} Saxton, C.~J., Soria, R., Wu, K., \& Kuin, N.~P.~M.\ 2012, \mnras, 422, 1625 

\bibitem[Saxton et al.(2017)]{saxton17} Saxton, R.~D., Read, A.~M., Komossa, S., et al.\ 2017, \aap, 598, A29 



\bibitem[Shiokawa et al.(2015)]{2015ApJ...804...85S} Shiokawa, H., Krolik, J.~H., Cheng, R.~M., Piran, T., \& Noble, S.~C.\ 2015, \apj, 804, 85 


\bibitem[Stone \& Loeb(2012)]{stone12} Stone, N., \& Loeb, A.\ 2012, \PRL, 108, 61302.

\bibitem[Stone et al.(2018)]{2018arXiv180110180S} Stone, N.~C., Kesden, M., Cheng, R.~M., \& van Velzen, S.\ 2018, arXiv:1801.10180 

\bibitem[Strubbe \& Quataert(2009)]{strubbe09} Strubbe, L.~E., \& Quataert, E.\ 2009, \mnras, 400, 2070 

\bibitem[Strubbe \& Murray(2015)]{strubbe15} Strubbe, L.~E., \& Murray, N.\ 2015, \mnras, 454, 2321 

\bibitem[Tchekhovskoy et al.(2011)]{tchekhovskoy11} Tchekhovskoy, A., Narayan, R., \& McKinney, J.~C.\ 2011, \mnras, 418, L79 

\bibitem[Tchekhovskoy et al.(2014)]{tchekhovskoy14} Tchekhovskoy, A., Metzger, B.~D., Giannios, D., \& Kelley, L.~Z.\ 2014, 
\mnras, 437, 2744 

\bibitem[Uhm(2011)]{uhm11} Uhm, Z.~L.\ 2011, \apj, 733, 86 

\bibitem[Ulmer(1999)]{ulmer99} Ulmer, A.\ 1999, \apj, 514, 180 

\bibitem[van Velzen et al.(2011)]{vanvelzen11} van Velzen, S., K{\"o}rding, E., \& Falcke, H.\ 2011, \mnras, 417, L51 

\bibitem[van Velzen et al.(2013)]{vanvelzen13} van Velzen, S., Frail, D.~A., K{\"o}rding, E., \& Falcke, H.\ 2013, \aap, 552, A5 

\bibitem[van Velzen et al.(2016a)]{vanvelzen16} van Velzen, S., Anderson, G.~E., Stone, N.~C., et al.\ 2016a, Science, 351, 62 

\bibitem[van Velzen et al.(2016b)]{2016ApJ...829...19V} van Velzen, S., Mendez, A.~J., Krolik, J.~H., \& Gorjian, V.\ 2016b, \apj, 829, 19 

\bibitem[\protect\citeauthoryear{van Velzen}{2018}]{2018ApJ...852...72V} van Velzen S., 2018, ApJ, 852, 72

\bibitem[van Velzen et al.(2018)]{2018arXiv180902608V} van Velzen, S., Gezari, S., Cenko, S.~B., et al.\ 2018, arXiv:1809.02608 

\bibitem[van Velzen et al.(2018b)]{vanvelzen18} van Velzen, S., Bower, G.~C., \& Metzger, B.~D.\ 2018, arXiv:1810.06677 

\bibitem[Vink{\'o} et al.(2015)]{2015ApJ...798...12V} Vink{\'o}, J.,  Yuan, F., Quimby, R.~M., et al.\ 2015, \apj, 798, 12 


\bibitem[Wang \& Liu(2016)]{wang16} Wang, X.-Y., \& Liu, R.-Y.\ 2016, \prd, 93, 083005 

\bibitem[Wang et al.(2018)]{2018MNRAS.477.2943W} Wang, T., Yan, L., Dou, L., et al.\ 2018, \mnras, 477, 2943 

\bibitem[Wiersema et al.(2012)]{wiersema12} Wiersema, K., van der Horst, A.~J., Levan, A.~J., et al.\ 2012, \mnras, 421, 1942 

\bibitem[Yang et al.(2016)]{yang16} Yang, J., Paragi, Z., van der Horst, A.~J., et al.\ 2016, \mnras, 462, L66 

\bibitem[Yuan et al.(2015)]{2015arXiv150607735Y} Yuan, W., Zhang, C., Feng, H., et al.\ 2015, arXiv:1506.07735 

\bibitem[Yuan et al.(2016a)]{yuan16} Yuan, Q., Wang, Q.~D., Lei, W.-H., Gao, H., \& Zhang, B.\ 2016a, \mnras, 461, 3375 

\bibitem[Yuan et al.(2016b)]{2016ApJ...828...92Y} Yuan, Y., Nalewajko, K., Zrake, J., East, W.~E., \& Blandford, R.~D.\ 2016b, \apj, 828, 92 

\bibitem[Zauderer et al.(2011)]{zauderer11} Zauderer, B.~A., Berger, E., Soderberg, A.~M., et al.\ 2011, \nat, 476, 425.

\bibitem[Zauderer et al.(2013)]{zauderer13} Zauderer, B.~A., Berger, E., Margutti, R., et al.\ 2013, \apj, 767, 152 


\end{thebibliography}
\end{document}